\documentclass[twocolumn,twocolappendix,trackchanges]{aastex701}
\usepackage{booktabs}
\usepackage{caption}
\usepackage{makecell}
\usepackage{amsmath}
\usepackage{ragged2e}
\usepackage{multirow}
\usepackage[title]{appendix}
\newcommand{\midsepremove}{\aboverulesep = 0mm \belowrulesep = 0mm}
\newcommand{\midsepdefault}{\aboverulesep = 0.605mm \belowrulesep = 0.984mm}

\begin{document}

\title{Kes 75 with IXPE: Detection of Nebular X-ray Polarization and Change in Pulsar Lightcurve}

\author[0000-0001-6395-2066]{Josephine Wong}
\email{joswong@stanford.edu}
\affiliation{Department of Physics and Kavli Institute for Particle Astrophysics and Cosmology, Stanford University, Stanford, California 94305, USA}

\author[0000-0002-6401-778X]{Jack T. Dinsmore}
\email{jtd@stanford.edu}
\affiliation{Department of Physics and Kavli Institute for Particle Astrophysics and Cosmology, Stanford University, Stanford, California 94305, USA}

\author[0000-0001-6711-3286]{Roger W. Romani}
\email{rwr@stanford.edu}
\affiliation{Department of Physics and Kavli Institute for Particle Astrophysics and Cosmology, Stanford University, Stanford, California 94305, USA}

\author[0000-0002-8665-0105]{Stefano Silvestri}
\email{stefano.silvestri@pi.infn.it}
\affiliation{Istituto Nazionale di Fisica Nucleare, Sezione di Pisa}

\author[0000-0002-2096-6051]{Shumeng Zhang}
\email{shumeng_zhang@connect.hku.hk}
\affiliation{Department of Physics, The University of Hong Kong, Pokfulam, Hong Kong}
\affiliation{Hong Kong Institute for Astronomy and Astrophysics, The University of Hong Kong, Pokfulam, Hong Kong}

\author[0000-0002-9652-0056]{Ruolan Jin}
\email{ruolanjin@gmail.com}
\affiliation{Department of Applied Science, National Taitung University, Taitung, Taiwan}

\author[0000-0002-4576-9337]{Matteo Bachetti}
\email{matteo.bachetti@inaf.it}
\affiliation{INAF Osservatorio Astronomico di Cagliari, Via della Scienza 5, 09047 Selargius (CA), Italy}

\author[0000-0002-5847-2612]{C.-Y. Ng}
\email{ncy@astro.physics.hku.hk}
\affiliation{Department of Physics, The University of Hong Kong, Pokfulam, Hong Kong}
\affiliation{Hong Kong Institute for Astronomy and Astrophysics, The University of Hong Kong, Pokfulam, Hong Kong}

\author[0000-0002-7574-1298]{Niccol\'{o} Di Lalla}
\email{niccolo.dilalla@stanford.edu}
\affiliation{Department of Physics and Kavli Institute for Particle Astrophysics and Cosmology, Stanford University, Stanford, California 94305, USA}

\author[0000-0002-9370-4079]{Wei Deng}
\email{dengw@st.gxu.edu.cn}
\affiliation{Guangxi Key Laboratory for Relativistic Astrophysics, School of Physical Science and Technology, Guangxi University, Nanning 530004, China}

\author[0000-0002-0105-5826]{Fei Xie}
\email{fei.xie@inaf.it}
\affiliation{Guangxi Key Laboratory for Relativistic Astrophysics, School of Physical Science and Technology, Guangxi University, Nanning 530004, China}
\affiliation{INAF Istituto di Astrofisica e Planetologia Spaziali, Via del Fosso del Cavaliere 100, 00133 Roma, Italy}

\author[0000-0002-6986-6756]{Patrick Slane}
\email{slane@cfa.harvard.edu}
\affiliation{Center for Astrophysics | Harvard \& Smithsonian, 60 Garden Street, Cambridge, MA 02138, USA}

\author[0000-0002-8848-1392]{Niccol\'{o} Bucciantini}
\email{niccolo.bucciantini@inaf.it}
\affiliation{INAF Osservatorio Astrofisico di Arcetri, Largo Enrico Fermi 5, 50125 Firenze, Italy}
\affiliation{Dipartimento di Fisica e Astronomia, Universit\`{a} degli Studi di Firenze, Via Sansone 1, 50019 Sesto Fiorentino (FI), Italy}
\affiliation{Istituto Nazionale di Fisica Nucleare, Sezione di Firenze, Via Sansone 1, 50019 Sesto Fiorentino (FI), Italy}

\author[0000-0002-3638-0637]{Philip Kaaret}
\email{philip.kaaret@nasa.gov}
    \affiliation{NASA Marshall Space Flight Center, Huntsville, AL 35812, USA}

\author[0000-0001-7263-0296]{Tsunefumi Mizuno}
\email{mizuno@astro.hiroshima-u.ac.jp}
\affiliation{Hiroshima Astrophysical Science Center, Hiroshima University, 1-3-1 Kagamiyama, Higashi-Hiroshima, Hiroshima 739-8526, Japan}

\author[0000-0001-7397-8091]{Maura Pilia}
\email{maura.pilia@inaf.it}
\affiliation{INAF Osservatorio Astronomico di Cagliari, Via della Scienza 5, 09047 Selargius (CA), Italy}

\author[0000-0001-9108-573X]{Yi-Jung Yang}
\email{yjyang312@gmail.com}
\affiliation{Center for Astrophysics and Space Science (CASS), New York University Abu Dhabi, PO Box 129188, Abu Dhabi, UAE}

\author[0000-0001-5326-880X]{Silvia Zane}
\email{s.zane@ucl.ac.uk}
\affiliation{Mullard Space Science Laboratory, University College London, Holmbury St Mary, Dorking, Surrey RH5 6NT, UK}

\author[0000-0002-5270-4240]{Martin C. Weisskopf}
\email{martin.c.weisskopf@nasa.gov}
\affiliation{NASA Marshall Space Flight Center, Huntsville, AL 35812, USA}

\begin{abstract}
We present the first X-ray polarization measurements of the PSR/PWN complex within SNR Kes 75. Two $\rm {\sim}\,500\,ks$ IXPE observations were conducted in October/November 2024 and April 2025. The second observation yields a significant phase-average 2--8 keV polarization degree $\rm PD=9.9\%\pm 2.5\%$ at $\rm PA=36.8^\circ\pm 7.3^\circ$, implying a toroidal field aligned with the PWN symmetry axis. The first epoch, however, has only a polarization upper limit. During this epoch, an additional pulsed component is visible at $\Delta \phi\approx 0.5$, detected at ${\sim}\,3.7\sigma$. An unbinned phase-resolved analysis reveals a high-PD rotating vector model PA sweep at the ${\sim}\,99.5\%$ confidence level, with angles fixed at those inferred from the PWN morphology; this can explain the loss of phase-average polarization. Additional observations are needed to pin down the nature of the anomalous pulse.
\end{abstract}


\section{Introduction} \label{sec:intro}

SNR G29.7-0.3 (Kes 75) is a Galactic supernova remnant and hosts the youngest known pulsar wind nebula (PWN) \citep{Reynolds2018}, powered by high field pulsar, PSR J1846-0258. It is $\rm {\sim}\,5.8\,kpc$ away \citep{Verbiest2012} with angular diameter $\rm d=3.5'$ \citep{Gotthelf2000} and ${\sim}\,25''\times20''$ PWN \citep{Guest2020}. The PWN has a rich morphology, with a compact nebula, southern jet, northern and southern clumps, and an arc-like ridge north of the pulsar, interpreted as a wisp or a part of the equatorial torus \citep{Ng2008}.

PSR J1846-0258 is a young ($\tau_c\,{\sim}\,700\,\rm{yrs}$), 330-ms pulsar with surface dipole field $\rm B_{surf}\,{\sim}\,5\times 10^{13}\,G$ and spin-down luminosity ${\sim}\,8\times10^{36}\,\rm{erg\,s^{-1}}$ \citep{Gotthelf2000}. It exhibited magnetar-like outbursts in 2006 and 2020, during which the 2.5--10 keV pulsed flux increased as much as ${\sim}\,10\times$ \citep{Gavriil2008,Hu2023} and a blackbody component ($kT\,{\sim}\,0.7\,\rm keV$) emerged \citep{Blumer2021}. The braking index permanently changed after the 2006 outburst, decreasing from $\rm n=2.65$ to $\rm n=2.16$, \citep{Livingstone2011}, and in both cases, the timing noise increased significantly and a spin-up glitch $\rm \Delta\nu = 2{-}4\times10^{-6}\,Hz/s$ was measured \citep{Kuiper2009,Hu2023}. Morphological changes were observed in the nebula as well: the southern inner jet doubled in flux and the northern clump broadened and shifted outwards by $2''$ to form a double-peaked feature \citep{Ng2008}. However, the integrated PWN flux did not change significantly \citep{Ng2008, Blumer2021}. To date, the only other rotation-powered pulsar that has shown magnetar-like outbursts is PSR J1119-6127, which is similarly young ($\tau_c<2\rm\,{kyr}$) and has high $\rm B_{surf}\,{\sim}\,4\times10^{13}\,G$ \citep{Archibald2016}. 

PSR J1846-0258 is undetected in the radio, with only upper limits for its average and single-pulsed emission (i.e. $\rm {<}\,8.2\ \mu Jy$ and $\rm {<}\,121.4\,\mu Jy$, respectively, at 1.25 GHz \citep{Juntao2025}, see also \cite{Sathyaprakash2024} for upper-limits at other bands from 700 MHz -- 13.9 GHz). However, the nebula is radio-bright, and 21 cm polarization maps show an integrated polarization $\rm PD = 10\%$ and $\rm PA = 80^\circ$ in a $\rm 10''$ region about the peak emission \citep{Becker1983}.  

To date, the Imaging X-ray Polarimetry Explorer (IXPE) \citep{Weisskopf2022} has measured the X-ray polarization of seven PWNe and five magnetars. In the PWNe, high polarization degrees, in some cases approaching the theoretical limit for synchrotron radiation \citep{Xie2022}, confirm the synchrotron origin of the X-ray emission. Moreover, for a majority of the PWNe, the integrated polarization angle is roughly aligned with the torus symmetry axis, suggesting nebular toroidal magnetic fields. Pulsar polarization has also been detected in a few cases, with the Crab showing an S-shaped PA sweep in the main pulse \citep{Wong2024}. 

Magnetar X-ray emission, on the other hand, is typically attributed to thermal hot spots from the stellar surface that may be re-polarized via resonant Compton scattering (RCS) off magnetospheric currents. Polarization is a useful tool to determine the emission mechanism, varying from very high for a magnetized atmosphere ($70-80\%$), to moderate for RCS (${\sim}\,30\%$), to very low for a condensed surface (${\lesssim}\,20\%$) \citep{Gonzalez2016, Taverna2020}. The five IXPE magnetar studies invoke all of these emission models (see \cite{Turolla2024} for review), indicating the wide variation of emission properties.

\begin{figure*}
\centering
\includegraphics[width=\linewidth]{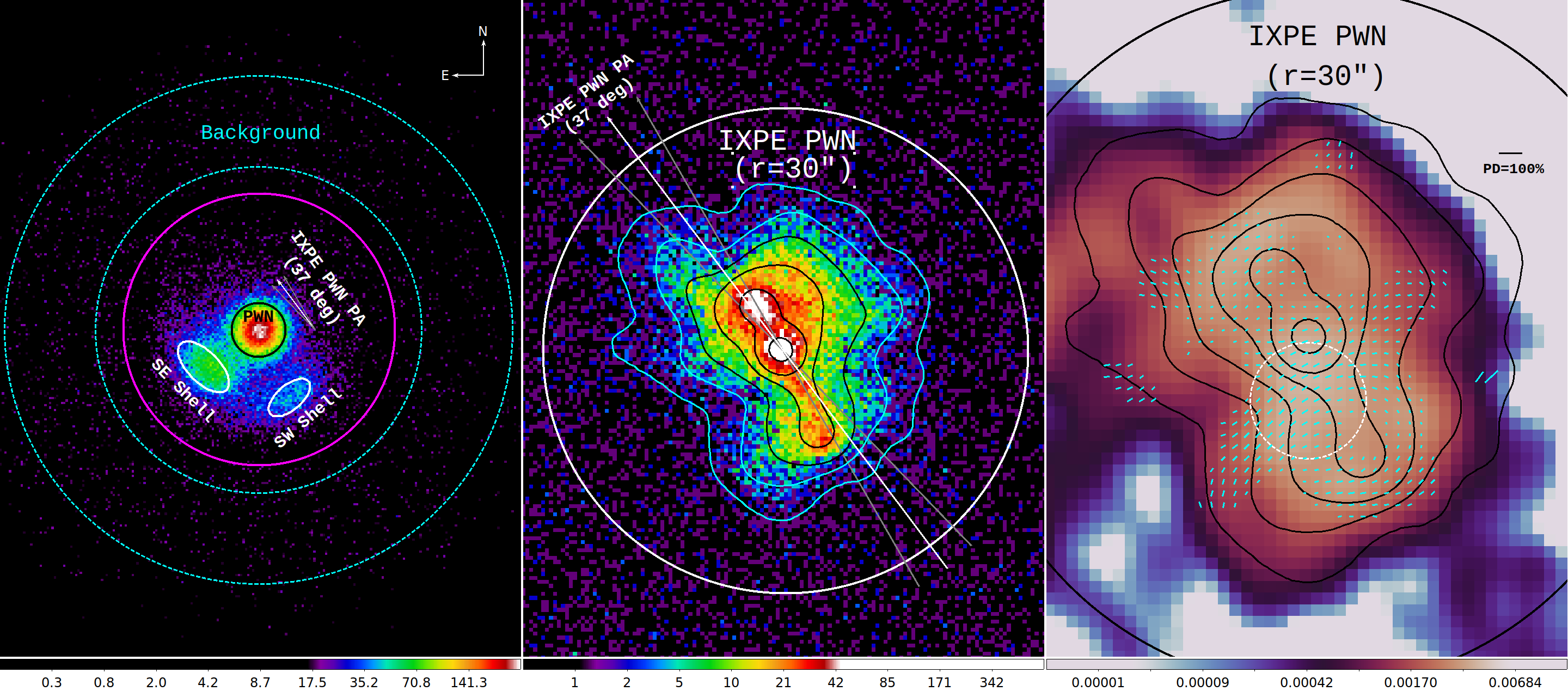}
\caption{IXPE (left), Chandra (middle), and eVLA (right) observations of Kes 75. \textbf{IXPE}: 2--8 keV counts, combining all detectors and observations, overlaid with PWN, PWN+SNR, SNR shells, and background extraction regions (the black PWN circle also represents IXPE's $\rm 30''$ resolution; refer to Section \ref{sec:pol} for exact radii specifications). The white (+gray) vector indicates the PWN polarization angle (+error) detected in the second IXPE observation. \textbf{Chandra}: Merged 2023 observations with intensity contours. \textbf{eVLA}: 2--4 GHz intensity map (Jy/beam) with ${>}\,3\sigma$ cyan polarization vectors rotated by $90^\circ$ to indicate magnetic field direction. Negative intensity values have been masked out. IXPE PWN and Chandra intensity contours included for reference. White dashed $\rm d=10''$ circle: here \cite{Becker1983} measured $\rm PD=10\%$ and $\rm PA=80^\circ$ at 21 cm.}
\label{fig:radio_xray}
\end{figure*}

In this paper, we report the first X-ray polarization measurements, along with contemporaneous spectral and timing properties, of Kes 75's PWN and PSR. Section \ref{sec:obs} describes the data reduction of IXPE, NICER, and archival Chandra observations. Section \ref{sec:pol} presents the phase-integrated polarization measurements using aperture photometry (hereafter `\texttt{PCUBE}') and spectro-polarimetric fitting. Sections \ref{sec:timing} and \ref{sec:phase_resolved} present the timing and phase-resolved polarization analysis, respectively. Section \ref{sec:discussion} discusses the results and Section \ref{sec:conclusion} concludes with a summary and future outlook.

\section{Observations and Data Reduction} \label{sec:obs}

IXPE observed Kes 75 between 31 October -- 11 November 2024 (Ep1) and 8 -- 19 April 2025 (Ep2), each epoch ${\sim}\,$500ks. NICER monitoring observations between 10 -- 13 November 2024 overlapped with the last two days of Ep1. Four 2023 ACIS-S Chandra observations were merged to create a flux model of the nebula and \cite{Gotthelf2021}'s pulsar spectroscopy was used to build a lightcurve model. The data reduction steps are described below. Unless otherwise specified, we have utilized \texttt{HEASOFT v6.34}, \texttt{IXPE ftools v1.9} and \texttt{ixpeobssim v31.1.0} \citep{Baldini2022}.

\subsection{IXPE}
The IXPE observations were obtained from the HEASARC archive\footnote{\url{https://heasarc.gsfc.nasa.gov/docs/ixpe/archive/}}. Ep1 was processed with an older pipeline with imperfect boom correction that left a trail of photons extending northwest of the PWN. Therefore, we reprocessed the event files with \texttt{HEASOFT v6.34}, the version used for Ep2, using a script provided by NASA SOC\footnote{This script is now publicly available at the Contributed IXPE Software webpage: \url{https://heasarc.gsfc.nasa.gov/docs/ixpe/analysis/contributed.html}}, significantly improving the boom correction. This is an important step since accurate position is needed to distinguish the PSR and PWN polarization components. The second observation was reprocessed with the same script for consistency. 

The standard reconstruction algorithm \texttt{ixpeevtrecon} relies on moments-weighting of the electron track to determine the photon interaction point and initial direction of the photoelectron, information used by downstream algorithms to determine the event position and polarization, as well as a set of quality weights. However, \cite{Peirson2021} developed a NN-based reconstruction algorithm that is able to provide better position reconstruction, azimuthal angle, and quality weighting. To improve our polarization measurements, we incorporate the NN-based reconstruction by running the standard pipeline up to \texttt{ixpegaincorrpkmap} (see the IXPE User Guide -- Software\footnote{\url{https://heasarc.gsfc.nasa.gov/docs/ixpe/analysis/IXPE-SOC-DOC-009D_UG-Software.pdf}} for a description of the workflow) and passing the output file to the NN reconstruction algorithm\footnote{\url{https://github.com/jack-dinsmore/ixpenn}}. It is updated with NN-reconstructed energies, positions, and quality weights and run through the rest of the standard pipeline. In all subsequent analysis, except for spectropolarimetric fitting, the NN-processed files are used. However, we also summarize the main results from moments-processed files, showing that they are consistent but at lower significance.

The Di Marco background rejection algorithm\footnote{\url{https://github.com/aledimarco/IXPE-background}} was used to remove particle background events. The image was shifted slightly between observations and detectors so we manually set the centroid position to the SIMBAD pulsar coordinates\footnote{\url{https://simbad.u-strasbg.fr/simbad/sim-id?Ident=PSR+J1846-0258}} ($18^h46^m24.75^s$, $-02^\circ58'29.6''$). These coordinates differ by ${\sim}\,3''$ from those reported in \cite{Helfand2003}, which can affect the barycentered times by up to ${\sim}\,10\,$ms. This should negligibly affect our results as it is only ${\sim}\,3$\% of J1846-0258's spin period. Photon arrival times were barycentered with \texttt{barycorr} using SIMBAD PSR coordinates and JPL--DE430 solar system ephemeris. Figure \ref{fig:radio_xray} shows the combined exposure 2--8 keV IXPE image.

An anomaly occurred in DU2 approximately 60\% through Ep2, where two rows of pixels failed and the detector gain increased. The IXPE SOC masked the bad pixels and re-calibrated the pulse intensities, but the Di Marco rejection algorithm still removed most post-anomaly DU2 events, as its filtering parameters were optimized using pre-anomaly data. To avoid this complication, we have excluded all post-anomaly DU2 events from our analysis, losing $\sim 13$\% of the total exposure.

In Appendix \ref{appendix:solar}, we describe our two methods for removing highly polarized solar background: (1) excising periods of solar flares with time cuts and (2) characterizing and subtracting the flares. They provide consistent polarization results, but the second method has smaller uncertainties (see Table \ref{tab:pcube}). 

\begin{table}[t]
\hspace{-2cm}
\begin{tabular}{ccccc}
\toprule[1.5pt]
\bf Obs ID & \bf MJD & \bf Date & \bf Livetime (s) & \bf Net Counts \\
\midrule
7033290113 & 60624 & 11/10/2024 & 413 & 1551 \\
7033290114 & 60625 & 11/11/2024 & 1376 & 3672 \\
7033290115 & 60626 & 11/12/2024 & 3906 & 10384 \\
7033290116 & 60627 & 11/13/2024 & 293 & 1136 \\
\bottomrule[1.25pt]
\end {tabular}
\captionof{table}{10 -- 13 November 2024 NICER observations. Note that the first two days overlap with Ep1. Livetime (seconds) and post-deflaring 2--8 keV counts are given.}\label{tab:nicer_obs}
\end{table}

\subsection{NICER}
NICER observations were processed with \texttt{nicerl2} in \texttt{HEASOFT v6.35.2}, utilizing both night and day settings to recover as many events as possible. Flares were filtered with \texttt{flaghighenergyflares}\footnote{\url{https://github.com/georgeyounes/NICERUTIL}} with probability threshold = 0.2. Events were barycentered in the same way as IXPE. Table \ref{tab:nicer_obs} describes the individual observations.

\subsection{Chandra}\label{sec:chandra}
A spatially-resolved nebula model was obtained using the most recent Chandra 2023 observations (ObsIDs 25211, 25735, 27771, 27772). The data were cleaned with \texttt{chandra\_repro} and aligned by shifting the brightest pixel to the SIMBAD PSR coordinates. The pulsar was removed by excising $\rm r < 1.5''$ and replacing it uniformly with events from $\rm 2'' < r < 3''$ at $1.35\times$ the annular count rate to produce a smooth image. Pileup is ${<}\,$1.5\% in the nebula. The four ACIS observations were merged with \texttt{reproject\_obs}.

\begin{figure*}
\centering
\includegraphics[width=\linewidth]{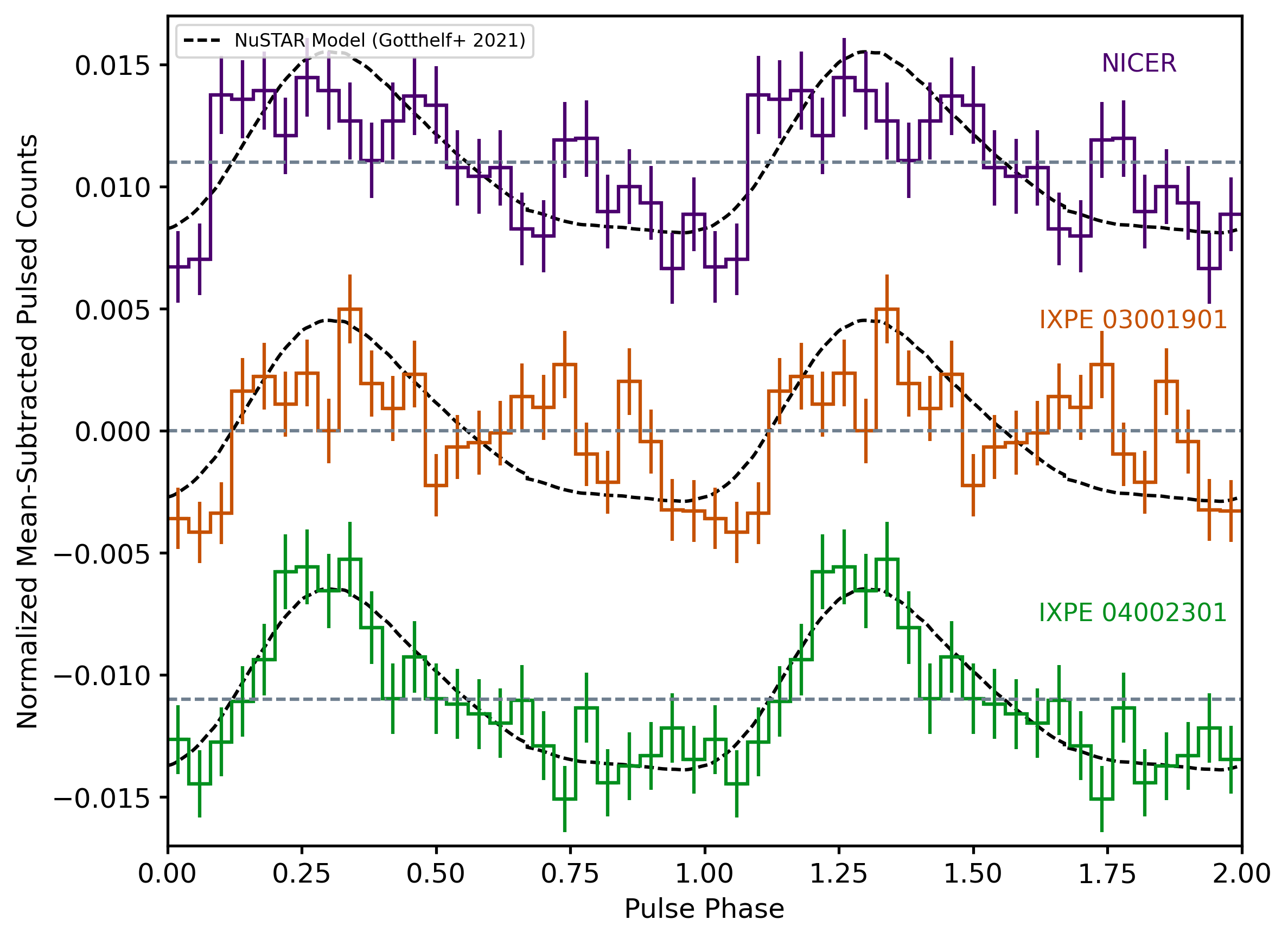}
\caption{IXPE and NICER binned pulse profiles using 2--8\,keV events $< 30''$ from the pulsar in the deflared dataset. Profiles are mean-subtracted and normalized by total counts. The NICER lightcurve (purple) is offset by +0.011 and the April 2025 (Ep2) IXPE lightcurve (green) is offset by $-0.011$. Error bars represent $1\sigma$ uncertainties. \cite{Gotthelf2021} model lightcurve derived from 2017 NuSTAR data is drawn for reference. Observations are aligned so that main pulse peak is located at $\phi=0.32$. \label{fig:lc_comparison}}
\end{figure*}

A pulsar lightcurve model was obtained by modulating the phase-averaged pulsar flux to match the pulsed profile and pulsed fraction measured by \cite{Gotthelf2021} using 2016 Chandra and 2017 XMM-Newton and NuSTAR observations. The pulsar's total spectrum was fitted with an absorbed power-law with $\Gamma=1.32$ and unabsorbed $\rm F_{\rm 2-10\ keV }=3.1\times10^{-12}\ erg\,cm^{-2}\,s^{-1}$. It has a 17\% pulsed fraction and a pulsed profile shown in their Figure 2. Using these values, and assuming constant $\Gamma$, we determined the phase-varying normalization that would produce the total pulsar flux with the correct modulation and pulsed fraction. Figure \ref{fig:lc_comparison} overlays our lightcurve model on the NICER and IXPE observations.

\section{Phase-Integrated Polarization}\label{sec:pol}

\subsection{\texttt{PCUBE}}
We analyzed the 2--8 keV integrated polarization in four regions: PWN ($\rm r<30''$), PWN + SNR ($\rm r<120''$), and the two SNR shells, where radii are expressed relative to the pulsar. These regions are drawn on the IXPE count map in Figure \ref{fig:radio_xray}. For our analysis, the Kislat prescription \citep{Kislat2015} with energy-dependent modulation factor and NN quality weights was used.

\begin{figure}
\includegraphics[width=\linewidth]{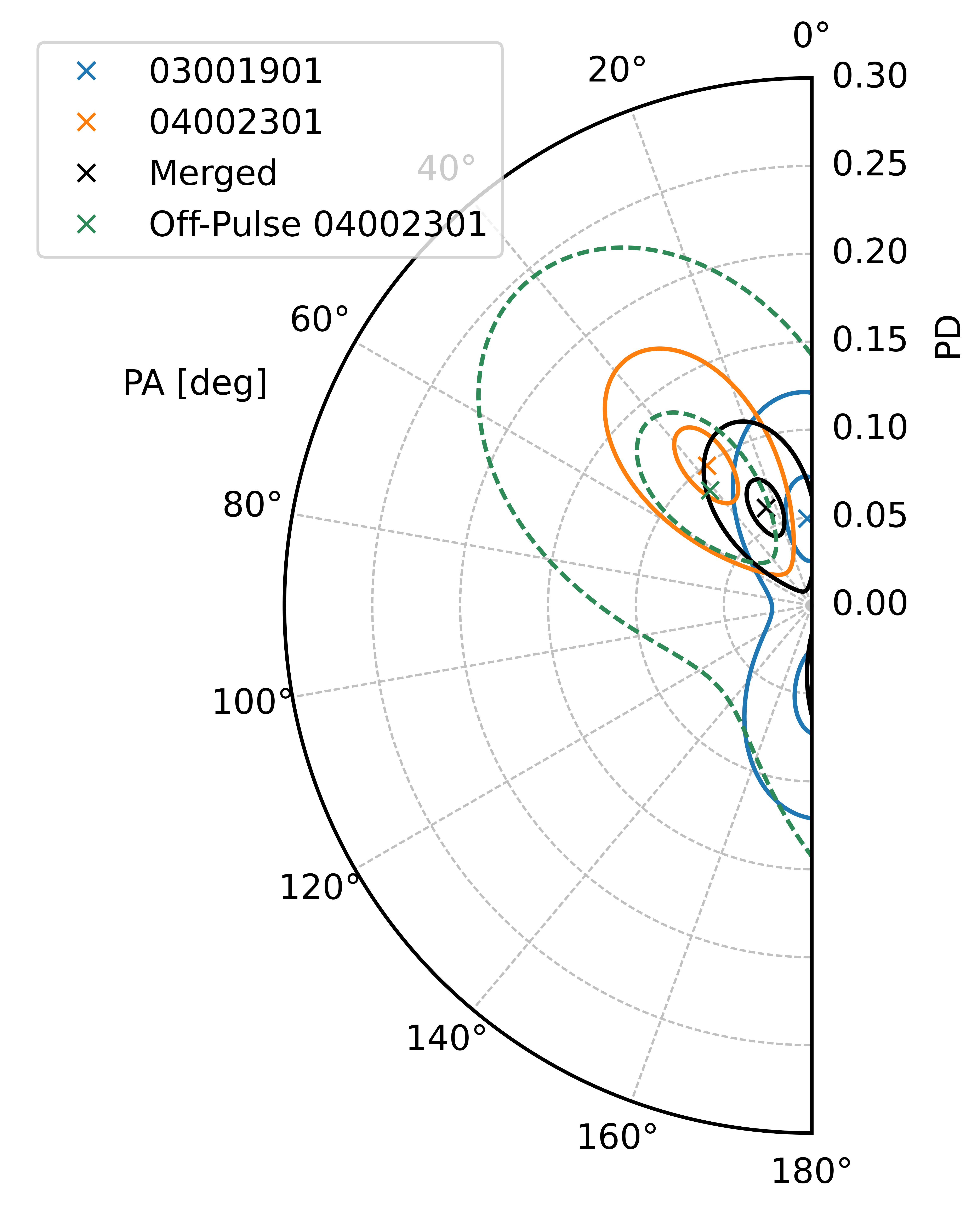}
\caption{2--8 keV PWN phase-integrated polarization for the Oct/Nov 2024 (Ep1), April 2025 (Ep2), and merged IXPE observations, and the off-pulse ($\Delta = 0.75-1$) Ep2 polarization. Inner/outer contours represent $1\sigma$/$3\sigma$ uncertainties, respectively.}
\label{fig:pwn_pd}
\end{figure}

Significant polarization is found for the PWN in Ep2 with $\rm PD=9.9\%\pm 2.5\%$ and $\rm PA=37^\circ \pm 7^\circ$. We do not, however, detect polarization in Ep1, with a $3\sigma$ upper limit at $\rm PD\,{\sim}\,12\%$. In the merged observation, PWN polarization is detected with $\rm PD = 6.1\% \pm 1.7\%$ and $\rm PA = 25^\circ\pm 8^\circ$, within $2\sigma$ of the Ep2 values, albeit at a lower $3.5\sigma$ significance. Table \ref{tab:pcube} summarizes the PWN results using the two methods of solar background removal. Figure \ref{fig:pwn_pd} shows the polarization contours of the individual and merged IXPE observations. We verify our results using moments-processed events and find that the Ep2 values are highly consistent, and the Ep1 values are still consistent within ${\sim}\,1.5\sigma$. 

\midsepremove
\begin{table*}
\centering
\hspace*{-2cm}
\begin{tabular}{c|ccc|ccc}
\toprule[1.25pt]\midrule
\multicolumn{7}{c}{\bf Method 1: Removing Solar Flares with Time Cuts} \\
\midrule
&  \multicolumn{3}{c|}{03001901}  &   \multicolumn{3}{c}{04002301} \\
\cline{2-7}
& Spectro-Pol (Mom) & \texttt{PCUBE} (Mom) &  \texttt{PCUBE} (NN) & Spectro-Pol (Mom) & \texttt{PCUBE} (Mom) & \texttt{PCUBE} (NN)  \\
\hline
$\rm\chi^2/DoF$           & 84.2/102           & ---                & ---               & 94.7/99           & ---                 & ---               \\
$\Gamma$                  & $1.94 \pm 0.03$    & ---                & ---               & $1.92 \pm 0.03$   & ---                 & ---               \\
$\mathcal{F}_x^\dagger$   & $1.29 \pm 0.09$    & ---                & ---               & $1.31 \pm 0.10$   & ---                 & ---               \\
Q/N                       & ---                & $0.058 \pm 0.031$  & $0.048 \pm 0.025$ & ---               & $0.014 \pm 0.034$   & $0.037 \pm 0.027$ \\
U/N                       & ---                & $0.059 \pm 0.031$  & $0.009 \pm 0.025$ & ---               & $0.092 \pm 0.034$   & $0.093 \pm 0.027$ \\
PD$^\lozenge$ (\%)        & $8.6 \pm 3.1$      & $8.2 \pm 3.1$      & $4.9 \pm 2.5$     & $9.2 \pm 3.4$     & $9.3 \pm 3.4$       & $10.1 \pm 2.7$    \\
PA$^\lozenge$ ($^\circ$)  & $23.7 \pm 10.5$    & $22.7 \pm 10.7$    & $5.31 \pm 14.6$   & $38.9 \pm 10.7$   & $40.7 \pm 10.4$     & $34.1 \pm 7.6$    \\
Significance              & 2.80               & 2.67               & 1.96              & 2.73              & 2.75                & 3.74              \\
\toprule[1.25pt]\midrule

\multicolumn{7}{c}{\bf Method 2: Characterizing and Subtracting Solar Background} \\
\midrule
&  \multicolumn{3}{c|}{03001901}  &   \multicolumn{3}{c}{04002301}  \\
\cline{2-7}
&   Spectro-Pol (Mom) &  \texttt{PCUBE} (Mom)  &  \texttt{PCUBE} (NN)  &   Spectro-Pol (Mom) &   \texttt{PCUBE} (Mom)   &  \texttt{PCUBE} (NN)  \\
\hline
$\rm\chi^2/DoF$           & 88.0/102          & ---                 & ---                  & 89.33/99         & ---                 & ---                 \\
$\Gamma$                  & $1.93 \pm 0.03$   & ---                 & ---                  & $1.92 \pm 0.03$  & ---                 & ---                 \\
$\mathcal{F}_x^\dagger$   & $1.30 \pm 0.09$   & ---                 & ---                  & $1.32 \pm 0.10$   & ---                 & ---                 \\
Q/N                       & ---               & $0.054 \pm 0.030$   & $0.049 \pm 0.024$    & ---              & $0.009 \pm 0.032$   & $0.028 \pm 0.025$   \\
U/N                       & ---               & $0.053 \pm 0.030$   & $0.005 \pm 0.024$    & ---              & $0.085 \pm 0.032$   & $0.095 \pm 0.025$   \\
PD$^\lozenge$ (\%)        & $8.0 \pm 3.0$     & $7.6 \pm 3.0$       & $4.9 \pm 2.4$        & $8.6 \pm 3.2$    & $8.5 \pm 3.2$       & $9.9 \pm 2.5$       \\
PA$^\lozenge$ ($^\circ$)  & $23.4 \pm 10.9$   & $22.3 \pm 11.2$     & $2.8 \pm 13.9$       & $39.4 \pm 11.0$   & $41.9 \pm 10.8$     & $36.8 \pm 7.3$      \\
Significance              & 2.70              & 2.56                & 2.06                 & 2.68             & 2.65                & 3.92                \\
\toprule[1.25pt]
\end{tabular}
\caption{2--8 keV PWN Polarization using \texttt{PCUBE} (Moments and NN-processed files) and Spectro-polarimetric Fitting (Moments only). All statistical uncertainties are quoted at the 68\% confidence level.
$^\dagger$Unabsorbed 2--8 keV flux ($\times10^{-11}$ erg cm$^{-2}$ s$^{-1}$). $^\lozenge$See Figure \ref{fig:pwn_pd} for full PD-PA uncertainty contours. Uncertainties for ${<}\,3\sigma$ measurements are reported here only as reference and do not reflect their full 2D covariance. \label{tab:pcube}}
\end{table*}
\midsepdefault

For comparison with X-rays, we analyze 2--4 GHz radio polarization obtained by the eVLA (Expanded Very Large Array). Figure \ref{fig:radio_xray} displays the radio intensity map overlaid with magnetic field vectors inferred from the radio polarization. The integrated polarization within the same $\rm r\,{<}\,30''$ PWN aperture is $\rm PD=14\%$ and $\rm PA=23^\circ$. Although \cite{Becker1983} report $\rm PD = 10\%$ with $\rm PA = 80^\circ$ in a $10''$ region at 21\,cm, our 2--4 GHz observation shows $\rm PD = 22\%$ and $\rm PA = 27^\circ$. 

The PWN-SNR complex is significantly detected in Ep2 with a lower $\rm PD=5.8\pm1.9\%$ and a consistent $\rm PA = 40^\circ\pm9^\circ$. The SW and SE shells are consistent with being unpolarized at $\lesssim 1\sigma$ for all observations.

\subsection{Spectro-Polarimetric Fitting}
We ran a spectro-polarimetric analysis of the PWN using moments-processed data to cross-check the \texttt{PCUBE} polarization. Unweighted Stokes I, Q, U spectra were extracted for each detector using \texttt{xpbin} in \texttt{ixpeobssim} and rebinned with \texttt{ftgrouppha} using the Kaastra and Bleeker optimal binning method to ensure Gaussian statistics. We used the \texttt{ixpeobssim} on-axis response functions with validity date 2024-07-01. The spectrum was modeled as an absorbed power-law with energy-independent polarization and an additional normalization to account for calibration differences between detectors: \texttt{constant $\times$ tbabs $\times$ polconst $\times$ powerlaw}. We fixed the absorption column $nH\rm=\,5.7 \times 10^{22}\, cm^{-2}$ \citep{Gotthelf2021,Blumer2021}, using \texttt{wilm} abundances and \texttt{vern} absorption cross-sections, as the absorption cannot be constrained in the IXPE energy range. The detectors were fitted jointly with all spectral and polarization parameters linked. 

As shown in Table \ref{tab:pcube}, this analysis confirms the \texttt{PCUBE} moments measurements at slightly higher significance. The 2--8 keV unabsorbed fluxes are not notably different between Ep1 and Ep2. They are also consistent with that measured by \cite{Blumer2021} during non-outburst states (PSR + PWN), after correcting to our extraction region and accounting for the observed secular flux decay. To this end, we simulated of the IXPE observation using \texttt{xpobssim} in \texttt{ixpeobssim}, convolving the merged 2023 Chandra observations (see Section \ref{sec:chandra}) with the IXPE instrumental response, and found that about ${\sim}\,20\%$ of the central PWN flux falls outside $\rm r = 30''$, averaged across detectors. Between 2000 and 2020, the nebular flux shows a linear decrease of ${\sim}\,0.45\%$ per year \citep{Blumer2021}. Accounting for these factors reproduces the flux measurements in \cite{Blumer2021} within ${\sim}\,2\sigma$ uncertainty.

\section{Timing Analysis} \label{sec:timing}

Since the latest published ephemeris \citep{Hu2023} only covers 8 August 2021 -- 14 November 2021, we used \texttt{HENzsearch} from the \texttt{HENDRICS} spectral-timing package to search for the pulsation about the extrapolated $\nu$ and $\dot{\nu}$. To minimize sky and nebular background, we used events within $\rm r = 30''$ of the pulsar from the deflared IXPE datasets. The oversampling rate was 32. First, we attempted a search with a single harmonic, but this yielded a non-detection for the first observation. Using two harmonics, a pulsation was detected at greater than 99.9\% confidence in both observations. The inferred $\dot{\nu} = -6.62 \times 10^{-11}$ Hz/s between the two IXPE epochs is similar to the individual epoch $\dot{\nu}$ found in the search, but uncertainties are too large to reliably phase connect. Table \ref{tab:ixpe_timing} reports the best ephemeris found for each observation. The Oct/Nov 2024 pulse is detected at $4.6\sigma$, while the April 2025 pulsation is detected at $6.4\sigma$. 

The two epochs appear to have different pulse profiles. Since there are four NICER epochs close to the 2024 IXPE observations, we have attempted to combine these for a comparison lightcurve, using only $\rm2-8\,keV$ events for equivalent comparison with IXPE. The individual NICER observations are too short to provide pulse detections, but a combined fold, searching near the extrapolated IXPE ephemeris, yields a $3.5\sigma$ pulsation, with ${\dot \nu}$ consistent with the IXPE measurements (Table \ref{tab:ixpe_timing}). In Figure \ref{fig:lc_comparison}, we compare the IXPE and NICER data with the historical lightcurve obtained by \cite{Gotthelf2021}. Lacking absolute phase, we have aligned the main peaks at $\phi=0.32$. IXPE Ep1 shows additional power near $\phi=0.8$, requiring a second harmonic in the pulse search. Although the signal-to-noise is not large, the NICER lightcurve also shows hints. Suggestion of a similar component in the 2018--2020 pre-outburst NICER profile can be seen in Figure 3 of \citet{Hu2023}, albeit primarily at lower energy. On the other hand, the Ep2 lightcurve matches well the historical pulse shape. 

To quantitatively determine the significance of the observed difference in the IXPE pulse profiles, we applied Kuiper's test, a variation of the Komolgorov-Smirnov (KS) test. Kuiper's test is invariant under cyclical transformations and is also sensitive to broader differences in the cumulative distribution than KS, which monitors only the extremum. The two-sample Kuiper's test yields a p-value of 0.0022 ($3.06\sigma$).

It is also useful to compare with the \cite{Gotthelf2021} pulse profile. Given the different resolutions and apertures, we must add an unpulsed $y$ component to this NuSTAR profile to model the IXPE data. The normalized probability distribution of observing a photon at phase $\{x\colon x\in[0,1)\}$ is then
\begin{equation}
g(x) = \frac{f(x) + y}{1+y}
\label{eq:pulse}
\end{equation}
\noindent where $f(x)$ is the model profile given by \cite{Gotthelf2021}, normalized to a phase-integrated flux of 1. For each IXPE epoch, $y$ was determined by maximizing the likelihood that $g(x)$ models the observed pulse profile. We utilized a uniform prior for $y$, employed 32 walkers and 5000 steps, removing a burn-in of 200 steps. We find $y=18^{+6}_{-4}$ for Ep1 and $y=10^{+2}_{-1}$ for Ep2, with errors marking the 16th and 84th percentiles of the posterior distribution. A similar exercise for the low statistics NICER profile finds $y=14^{+4}_{-3}$.

\begin{figure}
\centering
\includegraphics[width=\linewidth]{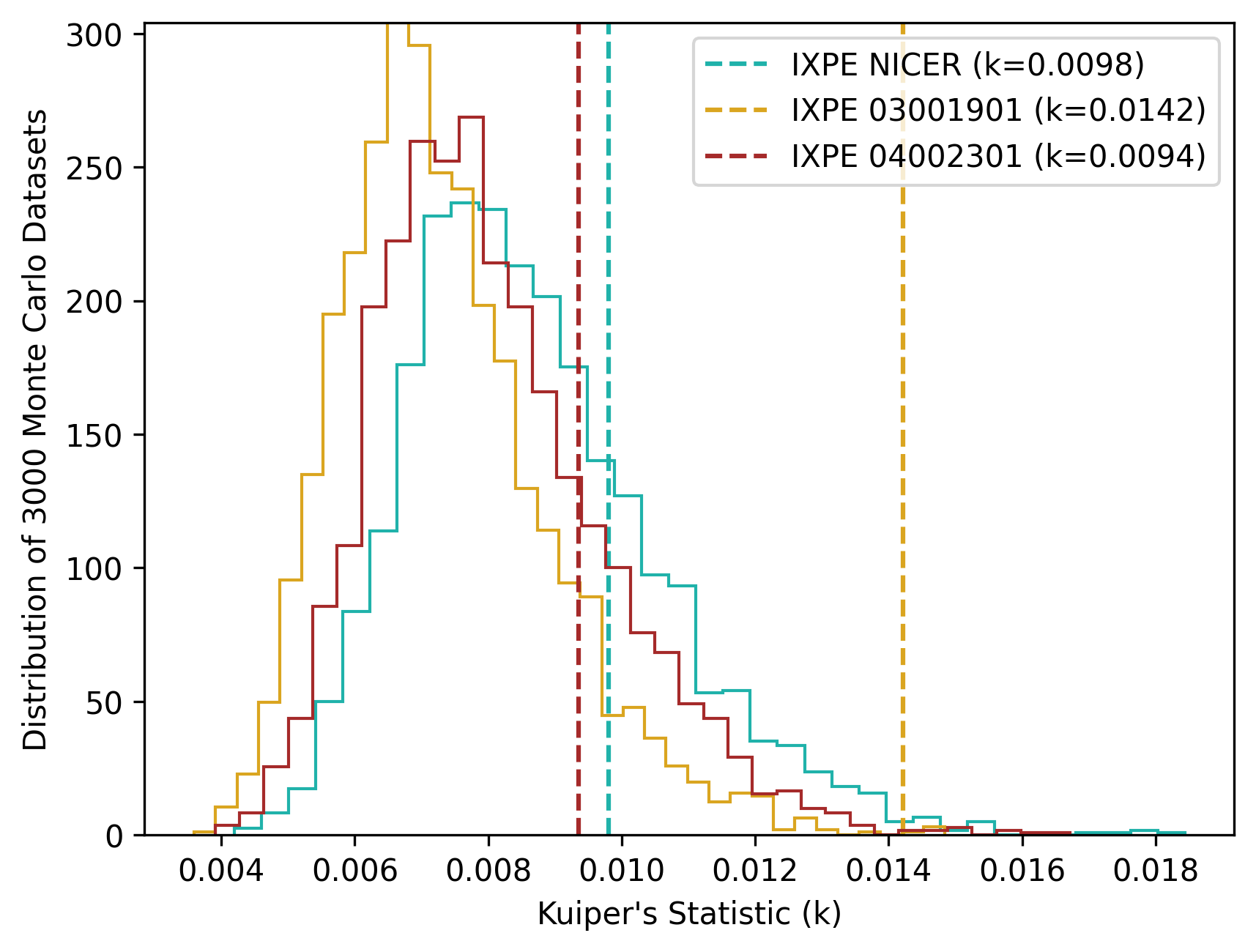}
\caption{Distribution of Kuiper's Statistic $k$ from 3000 Monte Carlo datasets, sampled from the best-fit pulse profile for NICER (aqua), Oct/Nov 2024 (Ep1, yellow), and April 2025 (Ep2, red) observations. Measured $k$ are indicated with the vertical dashed lines in the corresponding colors; these are at the 24.5th, 99.9th, and 20.5th percentiles, respectively. \label{fig:kuiper}}
\end{figure}

With $y$ determined, we can use Kuiper's test to determine how well the NuSTAR-derived model matches the observed lightcurves. To calculate the significance, we simulated 3000 random data sets from the $g(x)$ model with the same number of events as the observed (IXPE, NICER) datasets and re-fit $y$ to obtain an empirical distribution for the Kuiper's statistic $k$. Figure \ref{fig:kuiper} shows this distribution, alongside the measured $k$, for each dataset. We conclude that the Ep1 IXPE lightcurve is indeed inconsistent with the NuSTAR model (p-value of 0.0010; 3.29$\sigma$), while the Ep2 data are quite consistent (p-value of 0.79; 0.26$\sigma$). The NICER observation is consistent with the NuSTAR model, albeit at low significance (p-value of 0.75; 0.31$\sigma$). Thus the Ep1 IXPE lightcurve is distinctly different to that found in the NuSTAR analysis. By Ep2, the IXPE lightcurve had returned to its historical form. 

The 2024 IXPE/NICER profile appears to have an extra component versus the historical model. To quantify this, we added a circularly-symmetric Gaussian (a.k.a. von Mises) component $h(x\,\vert\,x_0,\kappa)={\rm{exp}}(\kappa\,{\rm{cos}}(x-x_0))/2\pi I_0(\kappa)$ to Equation \ref{eq:pulse}, which has three free parameters: location $x_0$, precision $\kappa$, and a relative amplitude $(1-k)$, giving the normalized distribution:
\begin{equation}
g\prime (x) = \frac{kf(x) + (1-k)h(x\,\vert\,x_0, \kappa) + y}{1+y}
\label{eq:pulse_gaussian}
\end{equation}

\begin{figure}
\centering
\includegraphics[width=\linewidth]{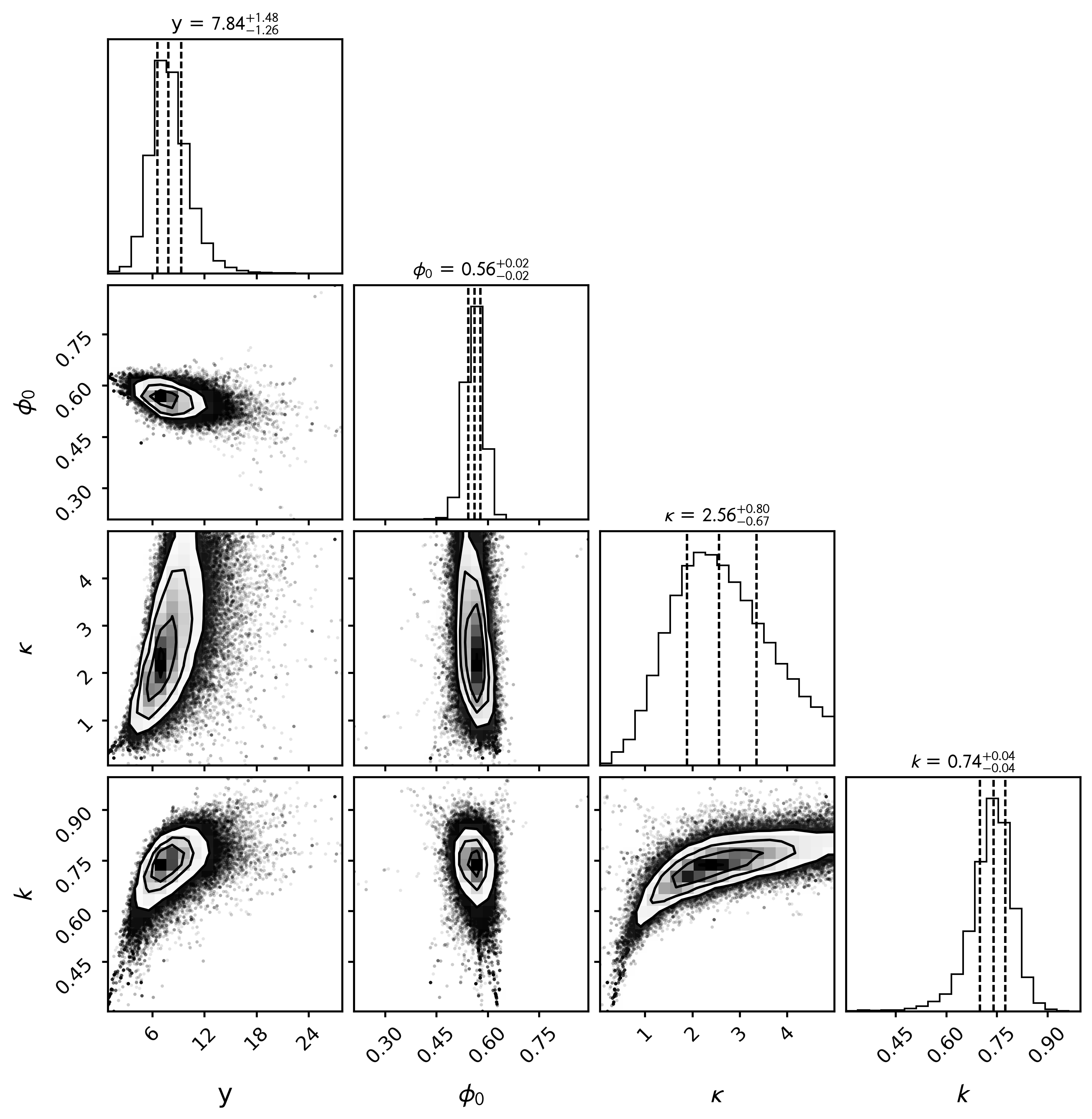}
\caption{Posterior Distributions of the two-pulse (with additional von Mises component) Ep1 lightcurve parameters. $y$ represents the DC component, $\phi_0$ ($x_0$ expressed in phase units) and $\kappa$ (1/rad$^2$) represent the location and precision of the second component, and $k$ is a scaling factor for the component flux ratio. Uncertainties are reported at the 16th and 84th percentiles. Data has been shifted by $\Delta\phi=-0.2$ so best-fit $\phi_0$ corresponds to $\phi=0.76$ in Figures \ref{fig:lc_comparison} and \ref{fig:phased_simple}.}
\label{fig:pulse_gaussian_fit}
\end{figure}

\noindent We again determine these parameters from a maximum-likelihood fit to the observed 2024 lightcurves. We ran 32 walkers for 5000 steps, removing a burn-in of 200 steps. The resulting posterior distributions for the four fit parameters for the 2024 IXPE data are shown in Figure \ref{fig:pulse_gaussian_fit}. 

Since Equation \ref{eq:pulse} is a nested version of Equation \ref{eq:pulse_gaussian}, a likelihood ratio test quantifies the improved match to the data. With 3 degrees of freedom, we obtain a p-value of $1.67\times10^{-4}$. Thus an extra von Mises component, roughly half a period offset and contributing $\sim 30\%$ of the pulse counts provides a good description of the 2024 lightcurves. 

The NICER observation lacks sufficient statistics to fit a second Gaussian component. In lieu, we compare the NICER observation with the two-component Ep1 profile and the single component Ep2 profile, and find a larger likelihood (13.8 vs 9.12) and a smaller Kuiper's statistic (0.011 vs 0.013) for the former comparison. Thus, the NICER 2024 data better matches IXPE Ep1 than the historical profile.

\textit{Moments-Processed Data:} To verify our results, we ran the same analysis with the moments-processed data. Fitting the historical profile to Ep1 and Ep2, we find $y=18_{-4}^{+6}$ and $y=9_{-1}^{+2}$, respectively. Ep1 is inconsistent with the NuSTAR model at a significance level of $p=0.0017$ ($3.14\sigma$) and Ep2 is consistent with $p=0.879$ ($0.15\sigma$). Fitting the two-pulse profile to Ep1, we obtain similar best-fit parameters: $y=6.84_{-1.59}^{+1.84}$, $\phi_0=0.58\pm0.2$, $\kappa=1.67_{-0.59}^{+0.81}$, and $k=0.70_{-0.08}^{+0.06}$ (see Figure \ref{fig:pulse_gaussian_fit} for the NN results). The likelihood ratio test has $p=1.78\times10^{-4}$, which suggests that the two-pulse model is a significantly better match. Comparison of the NICER data to the fitted Ep1 and Ep2 profiles is a little ambiguous. Kuiper's statistic is slightly smaller for Ep1 (0.012 vs 0.015), but the likelihoods are nearly the same (13.7 vs 13.8).

\begin{table*}
\begin{center}
\begin{tabular}{c c c c}
\hline\hline
Observation & NICER & IXPE 03001901 & IXPE 04002301 \\
\hline
Epoch ($t_0$, MJD) & 60625.82931 & 60619.54211 & 60778.74018 \\
$\rm \nu\ (Hz)$    & 3.028917576 & 3.028953805 & 3.028043244 \\
$\rm\dot{\nu}\ (10^{-11}\ Hz\ s^{-1})$ & -6.60 & -6.64 & -6.61 \\
$Z_2$ statistic    & 33.76 & 51.58 & 72.82 \\
p--value           & $4.27\times10^{-4}$ & $5.01 \times 10^{-6}$ & $1.66\times10^{-10}$ \\
\hline
\end{tabular}
\end{center}
\vspace{-4pt}
\caption{Highest $Z_2$-Statistics Ephemeris for PSR J1846-0258 during the NICER and IXPE Observations, determined by performing $Z_2$ search about the latest reported ephemeris for PSR J1846-0258 \citep{Hu2023}. Inferred $\dot{\nu}$ between observations is consistent with the best-fit $\dot{\nu}$ found through the search. Pulsation significance (expressed as p-value) accounts for the number of frequency samples used in the search. \label{tab:ixpe_timing}}
\end{table*}

\section{Phase-Resolved Polarization} \label{sec:phase_resolved}

\subsection{Binned Analysis}
\begin{figure}
\includegraphics[width=\linewidth]{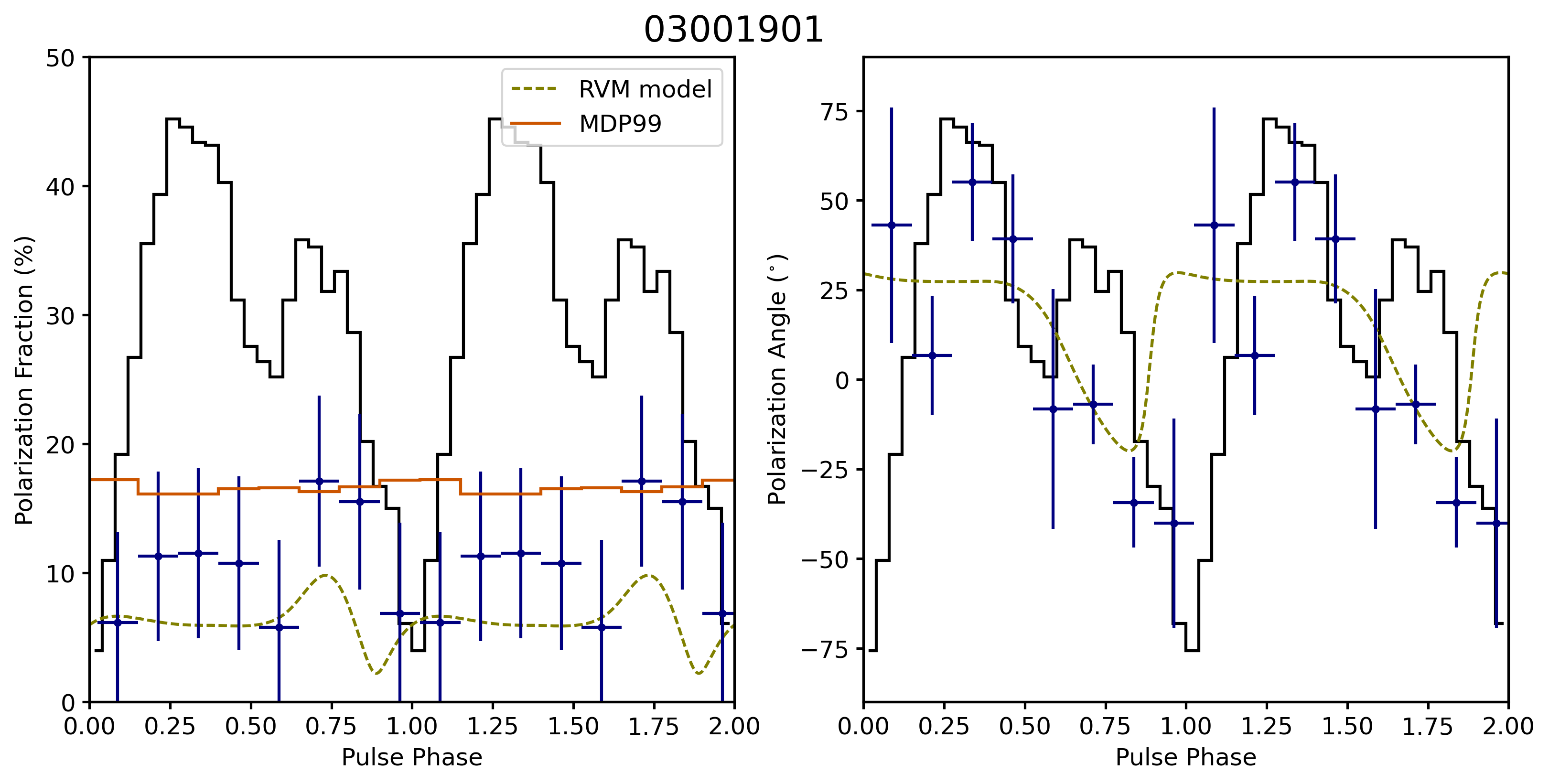}
\includegraphics[width=\linewidth]{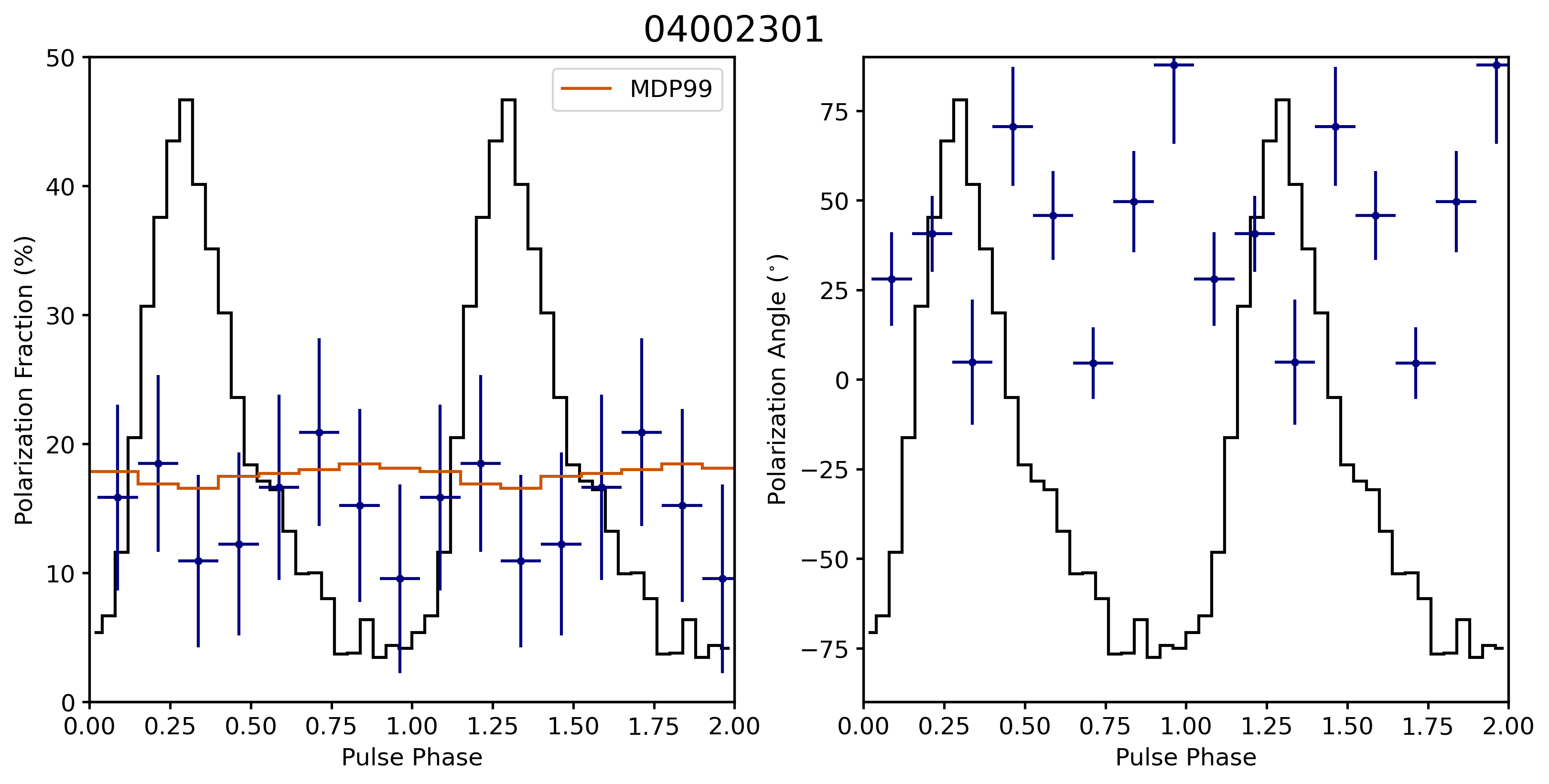}
\caption{Phase-binned polarization (8 independent bins) for the two IXPE observations. Only includes 2-8 keV events within $r < 30''$ from the pulsar, and no DC-subtraction has been performed. Best-fit RVM model (including contributions from nebula and both pulses, see Section \ref{sec:rvm_sec1}) for Ep1 (measured at ${\sim}\,99.5\%$ CL) is represented by dashed olive-green. 
No bins are significantly measured so PD-PA covariance is significant; the 1D error bars are plotted only for reference. $\rm MDP_{99}$ (orange) and smoothed lightcurve (black) lines are overlaid.\label{fig:phased_simple}}
\end{figure}

We divided the 2--8 keV, $\rm r\,{<}\,30''$ photons into eight phase bins and calculated the binned polarization for each IXPE observation (see Figure \ref{fig:phased_simple}). Note that this polarization is total, not nebula-subtracted. One Ep1 phase bin has $\rm PD\,{>}\,MDP_{99}$, with $\rm PD = 17.1\% \pm 6.6\%$ and $\rm PA = -6.9^\circ \pm 11.1^\circ$.  Two Ep2 bins are similarly significant: $\rm PD = 18.5\% \pm 6.8\%$ and $\rm PA = 40.6^\circ \pm 10.6^\circ$ for the earlier bin and $\rm PD = 20.9\% \pm 7.3\%$ and $\rm PA = 4.5^\circ \pm 10.0^\circ$ for the later one. The other phase bins for both observations show only $1-2\sigma$ $\rm PD$. 


The Ep2 lightcurve has an off-pulse region between $\phi = 0.75-1.00$. The Stokes parameters are $\rm Q/N = 0.011 \pm 0.053$ and $\rm U/N = 0.087 \pm 0.053$, or equivalently, $\rm PD = 8.7\%$ and $\rm PA = 41^\circ$ (see Figure \ref{fig:pwn_pd} for full 2D uncertainties), consistent ${<}\,1\sigma$ of the phase-integrated PWN polarization found in Section \ref{sec:pol}. The Ep1 observation has no apparent off-phase. If we attempt to use the Ep2 `off-pulse' interval, a very different $\rm PD=9.3\%$ and $\rm PA=-37^\circ$ is obtained. Thus, off-pulse subtraction to remove the nebular component will not work well for these data. 

\begin{figure}
\includegraphics[width=\linewidth]{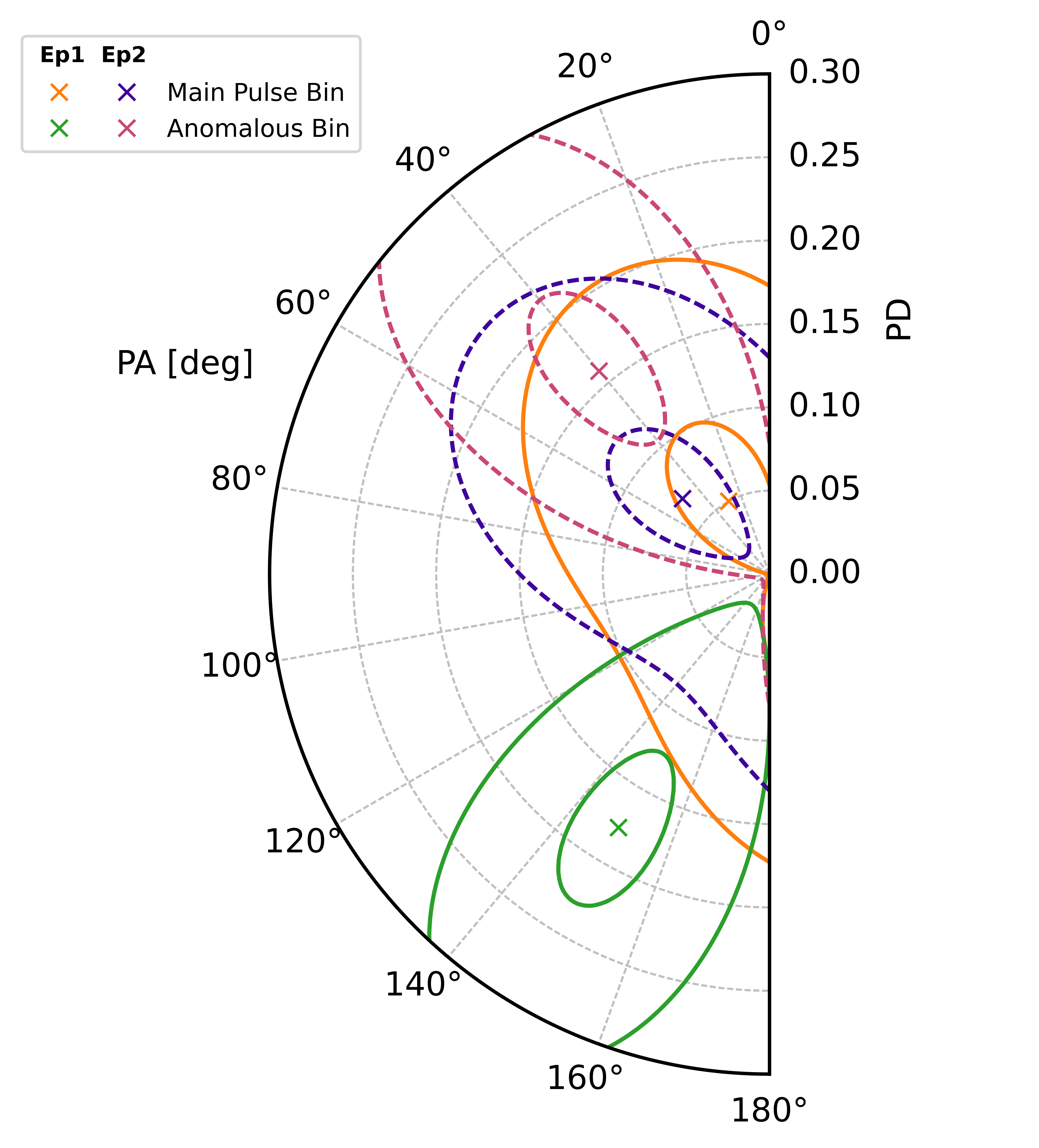}
\caption{1$\sigma$ and 3$\sigma$ polarization contours for coarse phase-binned analysis of Ep1 (orange, green) and Ep2 (purple, red). `Main Pulse Bin' and `Anomalous Bin' correspond to $\phi=(0, 0.5)$ and $\phi=(0.5,1)$ in Figure \ref{fig:phased_simple}. For Ep1, the two bins differ by ${>}\,3\sigma$, with the anomalous pulse detected at $\rm PD=17.7\%\,\pm\,5.2\%$ at a significant angle relative to the main pulse. Ep2 second bin is marginally detected above $\rm MDP_{99}$ at $\rm PD = 15.9\% \pm 5.5\%$ and $\rm PA=40.0^\circ\pm9.9^\circ$, and is consistent with the main pulse bins of Ep1 and Ep2 at ${\sim}\,2\sigma$.  \label{fig:phased_coarse}}
\end{figure}

We have also attempted a coarser analysis with a main pulse bin $\phi=(0, 0.5)$ and an anomalous pulse bin $\phi=(0.5, 1)$ (see Figure \ref{fig:phased_coarse}). In Ep1, we detect polarization in the anomalous pulse with $\rm PD=17.7\%\,\pm\,5.2\%$ and $\rm PA=149.2^\circ\pm 8.4^\circ$, differing by ${>}\,3\sigma$ from the main pulse. In Ep2, the second bin is marginally detected above $\rm MDP_{99}$ with $\rm PD = 15.9\% \pm 5.5\%$ and $\rm PA=40.0^\circ\pm9.9^\circ$ and agrees with the main pulse bin within $2\sigma$. To further study the pulsar polarization, we performed an unbinned phase-resolved polarization analysis.

\subsection{Polarization Sweep Modeling} \label{sec:rvm}

Figure \ref{fig:phased_simple} shows substantial differences of the phase-binned EVPA between the two observation, despite the low significances. In Ep1, the polarization angle appears to sweep ${
\sim}130^\circ$ over phase $\phi=0.3-1$. The low 10-20\% polarization degree in this range may in part be due to a rapid sweep, which, if real, must originate from the pulsar. In contrast, the Ep2 data show no organized phase structure. To further explore the apparent angle sweep we conduct an unbinned MCMC fit, using the rotating vector model (RVM).

\subsubsection{RVM Model}
The pulsar phase-resolved polarization angle is traditionally fit with the rotating vector model (RVM):
\begin{equation}
\begin{split}
\rm tan(&\psi - \psi_0) = \\
&\rm \frac{sin\,\theta\times sin(\phi-\phi_0)}{cos\ \zeta\times sin\ \theta\times cos\ (\phi-\phi_0)-sin\ \zeta\times cos\ \theta} \label{eq:rvm}
\end{split}
\end{equation}
with inclination angle $\zeta$, magnetic obliquity (angle between magnetic axis and spin axis) $\theta$, and the phase and polarization angle at the closest approach of the magnetic axis, $\phi_0$ and $\psi_0$. Their bounds are given by

\begin{equation*}
\begin{split}
\zeta &\in [-180^\circ, 180^\circ] \\
\theta & \in[-90^\circ,90^\circ] \\
\phi_0 &\in [-180^\circ, 180^\circ] \\
\psi_0 &\in [-90^\circ,90^\circ]
\end{split}
\end{equation*}

An unbinned fit was conducted with the same cuts as the binned analysis ($r<30''$ and 2--8 keV) using a Goodman-Weare Monte Carlo sampler. We assume two source components: an unpulsed component with phase-independent polarization ($p_{0,\,\rm neb}$, $\psi_{0,\,\rm neb}$) and a pulsed component with constant polarization degree ($p_{0,\,\rm psr}$) and polarization angle modeled by the RVM. The current data does not provide enough sensitivity to measure the pulsar PD phase dependence. Moreover, we assume only the anomalous pulse is polarized (meaning the main pulse is fixed at $\rm PD=0\%$, the data lacks sufficient statistics to measure its polarization). 

Note that the unpulsed component may have some contribution from pulsar DC flux. However, because it is predominantly nebular, we have decided to call this component, the `nebula.' Similarly, the RVM sweep only represents the pulsed component of the pulsar polarization, but for simplicity, we have decided to call the pulsed component, the `pulsar'. 

The likelihood function that we strive to maximize is given by:

\begin{equation}
\hspace{-1.5cm}
\begin{split}
\mathcal{L} = \prod_{i=1}^{N}\ & q_i \Bigl[z(\psi_i\,\vert \,\phi_i,\zeta,\theta,\phi_0,\psi_0,p_{0,\,\mathrm{psr}})P(\phi_i,x_i,y_i\,\vert\,{\mathrm{psr}})P({\mathrm{psr}})\\
&+z(\psi_i\,\vert\,p_{0,\,\mathrm{neb}},\psi_{0,\,\mathrm{neb}})P(\phi_i,x_i,y_i\,\vert\,{\mathrm{neb}})P({\mathrm{neb}})\Bigr]
\end{split}
\end{equation}

\noindent where $z(\psi)$ is the probability distribution of the measured azimuthal angle $\psi$ from a source with true polarization degree $p_0$ and expected polarization angle $\psi_0$, given by \cite{Kislat2015}, Eq. 8:

\begin{equation}
z(\psi)=\frac{1}{2\pi}(1+p_0\,\mu\,\rm{cos}[2(\psi-\psi_0)])
\label{eq:kislat}
\end{equation}

\noindent where $\mu$ is the modulation factor of our detector/analysis scheme. The probability $P({\phi_i,x_i,y_i\,\vert\,\mathrm{psr}})$ is given by
\begin{equation}
P({\phi_i,x_i,y_i\,\vert\,\mathrm{psr}}) = P(x_i,y_i \vert {\rm psr})\times P(\phi_i \vert {\rm psr})
\end{equation}

\noindent and similarly for the nebula, since the probabilities of detecting an event at ($x_i$, $y_i$) and at $\phi_i$ are independent. We used \texttt{Leakagelib} \citep{Dinsmore2025} functionalities to calculate $P(x_i,y_i\,\vert\,\rm psr)$, which utilizes non-axisymmetric sky-calibrated point spread functions (PSFs) \citep{Dinsmore2024} to construct the spatial distribution of the IXPE counts. $P(\rm psr)$ is the relative flux of the pulsar, such that $P({\rm psr}) + P({\rm neb})=1$. $q_i$ are the NN-derived quality weights \citep{Peirson2021}. We used 16 walkers per free parameter and 10000 steps, removing a burn-in of 1000 steps, to fit the model parameters.  

\subsubsection{RVM Fitting} \label{sec:rvm_sec1}
In the RVM, the EVPA is typically assumed to be parallel (curvature radiation) or perpendicular (synchrotron radiation) to the projected spin axis $\Psi$. \cite{Ng2008} measured a projected spin axis $\Psi = 207^\circ \pm 8^\circ$ along the southern jet and an inclination angle $\zeta = 62^\circ \pm 5^\circ$ (out of the plane of the sky). Equivalently, we can set the opposite pole as reference, i.e. $\Psi = 27^\circ$, $\zeta= 118^\circ$ (into the plane of the sky). 

Since the nebula symmetry (pulsar spin) axis is robustly measured, we begin by fixing $\Psi=27^\circ$ and assigning the EVPA relative to $\Psi$: $\psi_0=\Psi$ (curvature) or $\psi_0=\Psi+90^\circ$ (synchrotron). All other RVM parameters have uniform priors. $p_{0, \rm psr}$ was allowed to vary uniformly between 0 and 1. 1D Gaussian priors for $p_{0,\rm neb}$, $\psi_{0,\rm neb}$ were assigned using the Ep2 PWN \texttt{PCUBE} measurements with doubled uncertainty. 

The best-fit synchrotron and curvature models have similar likelihoods. In both cases, $\zeta$ shows modes around $\pm118^\circ$. However, a bright arc northeast of the pulsar, interpreted as the Doppler-boosted termination shock, prefers the $+118^\circ$ solution so we select for this mode. In addition, we find two $\theta$/$\phi_0$ modes, representing the two poles: $+35^\circ$/$1.0$ and $-35^\circ$/$0.5$ for the synchrotron case, and $-60^\circ$/$1.0$ and $+60^\circ$/$0.5$ in the curvature case. Note that the $\theta$/$\phi_0$ modes give the same RVM sweep and represent the same dipole geometry. In all scenarios, we report the mode that yields the largest likelihood. Table \ref{tab:rvm_unbinned} records the best-fit parameters for both scenarios.

To determine the significance of the best-fit models, we compare it with a constant PA model. We fix $p_{0, \rm psr} = 0\%$, $\theta=0^\circ$, and $\phi_0$, $\psi_0$, and $\zeta$ at arbitrary values. Since the constant model is nested within the RVM, we can utilize the likelihood ratio test to determine the significance, finding $p=0.037$ for the synchrotron model and $p=0.035$ for the curvature model.

Since the best-fit $\zeta$ is consistent with the externally-modeled value, we fix $\zeta = 118^\circ$ and re-fit the parameters. The best-fit parameters differ by at most 5\% from the free fit and the likelihoods increase by ${\lesssim}\,0.02$. The likelihood ratio test, with one fewer degree of freedom, now yields $p=0.019$ and $p=0.011$ for the synchrotron and curvature scenarios, respectively.

\begin{figure}
\centering
\includegraphics[width=\linewidth]{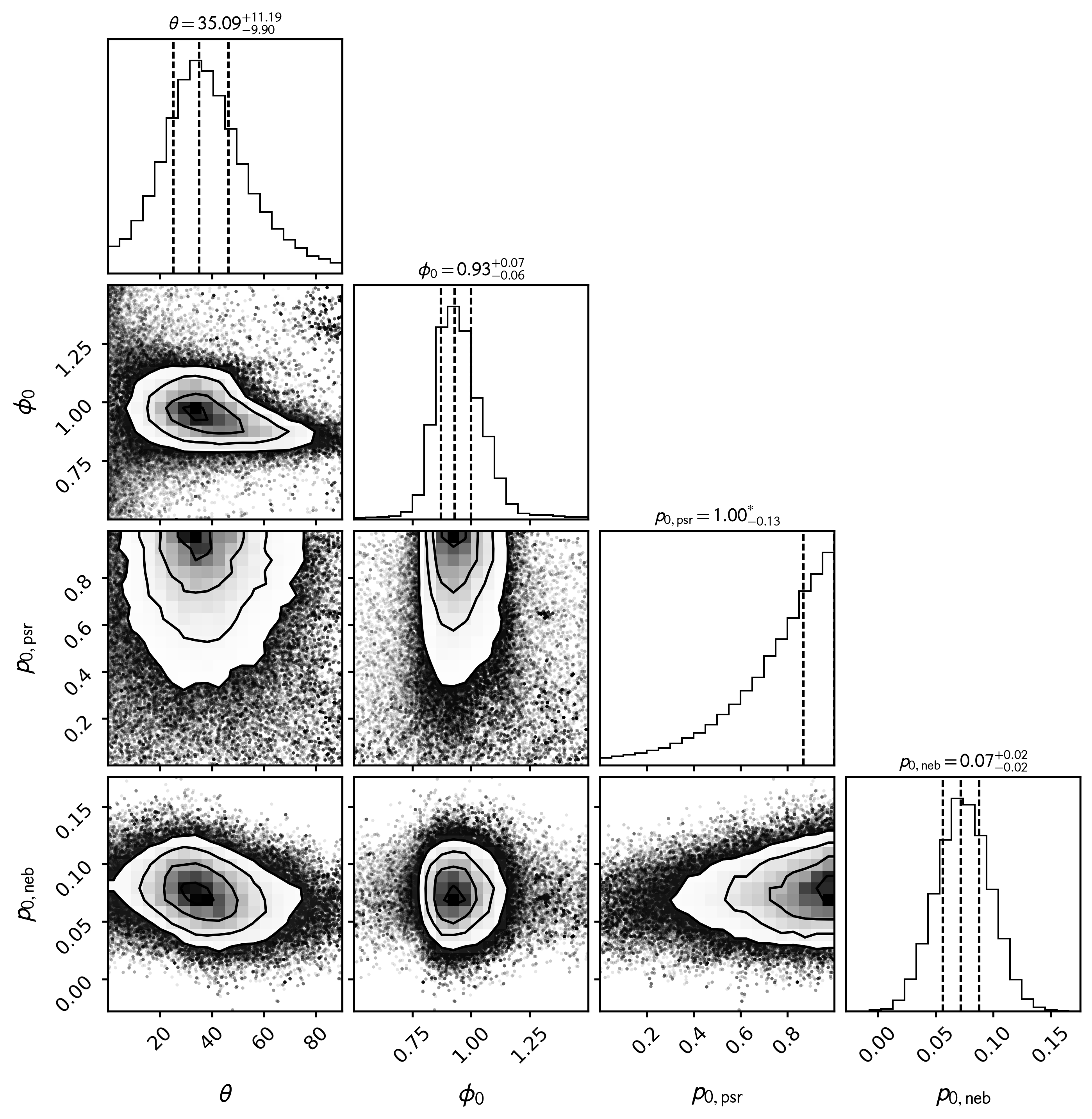}
\caption{\label{fig:rvm_unbinned} Posterior Distributions of IXPE Oct/Nov 2024 (Ep1) Best-Fit RVM Model. Pulsar has been modeled with constant PD and RVM PA with $\zeta=118^\circ$ and $\psi_{0,\rm psr}=117^\circ$ fixed. Nebula has been modeled with fixed $\psi_{0,\rm neb}=27^\circ$ and constant PD with a Gaussian prior based on the Ep2 \texttt{PCUBE} measurement (see Table \ref{tab:pcube}). The dominant mode ($\phi_0\,{\sim}\,1.0$) has been selected for, but a secondary mode exists at approximately $\phi_0 \rightarrow \phi_0+0.5$ and $\theta \rightarrow -\theta$ that produces the same sweep profile. All angle parameters are expressed in degrees and $\phi_0$ is expressed in phase units.}
\end{figure}

As seen in Table \ref{tab:rvm_unbinned}, the best-fit $\psi_{0,\rm neb}$ is always near the measured symmetry axis so we fix $\psi_{0,\rm neb} = 27^\circ$. The RVM models are essentially unaffected, but the constant model has a slightly smaller likelihood. The likelihood ratio test now yields $p=0.006$ and $p=0.009$, exceeding the 99\% confidence level. The Jeffreys scale $\rm exp[({AIC}_{const}-{AIC}_{unconst})/2]$, which assesses two models' AICs, yields 24 and 16, respectively, which ``strongly" favors the RVM model. Table \ref{tab:rvm_unbinned} reports the best-fit parameters. Figure \ref{fig:rvm_unbinned} shows the posterior distributions for the most significant (constrained synchrotron) model.

To compare with Ep1 PWN \texttt{PCUBE} measurement, we calculated the model phase-integrated polarization, finding $\rm PD\,{\sim}\,5\%$, $\rm PA\,{\sim}\,15^\circ$ for both scenarios, consistent within $1\sigma$ of the \texttt{PCUBE} measurement. Thus, an RVM-like polarized component can explain the low polarization found in Ep1.

Furthermore, we checked whether our best-fit model is consistent with the Ep2 data. We fit the same model -- a constant nebula polarization and an unpolarized main pulse -- to Ep2 and compare the likelihood of this model with the model obtained with Ep1. The best-fit parameters are $p_{0, \rm neb}=9.7\% \pm 1.8\%$ and $\psi_{0, \rm neb}=35^\circ \pm 5^\circ$. The highest-significance Ep1 RVM model (constrained synchrotron) has $p_{0, \rm neb} = 7.2\% \pm 1.6\%$ and $\psi_{0, \rm neb} = 27^\circ$. The likelihood ratio test yields $p=0.330$, meaning the two models equally well explain the data.

It is useful to note that the impact parameter $\beta = \vert \zeta -\theta\vert$ is large (${\sim}\,85^\circ$ for synchrotron model, ${\sim}\,60^\circ$ for curvature model), which means that our line-of-sight passes far from the magnetic axis. This is consistent with the non-detection of radio pulsations.

We have also tried fitting a totally unconstrained RVM model. The best-fit $\zeta=113^{+30}_{-35}$ deg is near the externally modeled value, although the uncertainty is quite large. The best-fit $\psi_0=150^{+40}_{-23}$ deg is different from that expected for either synchrotron or curvature radiation based on the projected spin axis measured by \cite{Ng2008}. The likelihood is marginally better (+0.5) than the fixed $\psi_0$ models, but with one additional degree of freedom, it has a lower significance $p=0.060$. This departure from the observed symmetry axis may be a noise fluctuation -- we should, however, thus avoid strong claims about the pulsar geometry based on our RVM fits.

\textit{Moments-Processed Data:} To check our results, we run the same analysis for the moments-processed dataset. With the totally unconstrained RVM model, we obtain $\zeta = 110_{-40}^{+36}$ deg, which is again close to the modeled value, and $\psi_0=146_{-28}^{+43}$ deg. If we constrain $\zeta$ and $\psi_{0, \rm neb}$, we find that all parameters are within a few percent of the parameters obtained with the NN-processed data, except for $p_{0, \rm neb}\,{\sim}\,10\%\pm2\%$, and the phase-integrated polarization is consistent within $1\sigma$ of the \texttt{PCUBE} measurements. However, the significances of the models are much smaller with $p\approx0.054$.  \\

\begin{table*}
\begin{center}
\hspace*{-3.7cm}
\begin{tabular}{c|c|c|c|c|cc}
\hline\hline
\multicolumn{1}{c|}{\multirow{2}{*}{Parameters}} & \multicolumn{2}{c|}{Synchrotron Model} & \multicolumn{2}{c|}{Curvature Model} & \multicolumn{2}{c}{Constant Model} \\ \cline{2-7}
\multicolumn{1}{c|}{} & \multicolumn{1}{c|}{Free} & \multicolumn{1}{c|}{Fixed $\zeta, \psi_{0, \rm neb}$} & \multicolumn{1}{c|}{Free} & Fixed $\zeta, \psi_{0, \rm neb}$ & Free & Fixed $\psi_{0, \rm neb}$ \\ \hline
$p_{0,\,{\rm{neb}}}$ (\%) & $6.7 \pm 1.6$ & $7.2 \pm 1.6$ & $6.5\pm 1.7$ & $6.6 \pm 1.5$ & $5.6 \pm 1.8$ & $5.6 \pm 1.6$ \\
$\psi_{0,\,{\rm{neb}}}$ (deg) & $29 \pm 6$ & 27 & $26 \pm 7$ & 27 & $16 \pm 9$ & 27 \\
$\zeta$ (deg) & $99_{-34}^{+32}$ & 118 & $121_{-35}^{+30}$ & 118 & --- & --- \\
$\theta$ (deg) & $33 \pm 13$ & $35 \pm 11$ & $-57 \pm 16$ & $-60 \pm 10$ & --- & --- \\
$\phi_0$ & $0.89 \pm 0.07$ & $0.93 \pm 0.07$ & $0.03 \pm 0.08$ & $0.01 \pm 0.07$ & --- & --- \\
LRT (p-value) & 0.037 & 0.006 & 0.035 & 0.009 & --- & --- \\
Integrated PD/PA & $4.2\%/18^\circ$ & $5.2\%/15^\circ$ & $5.5\%/14^\circ$ & $5.0\%/15^\circ$ & --- & ---\\       
\hline
\end{tabular}
\end{center}
\vspace{-4pt}
\caption{Best-Fit Models (including constant models) obtained from unbinned maximum-likelihood fit to the Oct/Nov 2024 (Ep1) IXPE observation. Only deflared events within 2--8 keV and $r < 30''$ circular aperture were included. $p_{0,\,{\rm{psr}}}$ has been assumed constant and we adopt $\Psi=27^\circ$ from the PWN morphology, with EVPA oriented parallel (curvature) for perpendicular (synchrotron) radiation. Best-fit values are the posterior modes, utilizing 16th/84th percentiles deviations from the median for uncertainties. In all models, $p_{0,\rm psr}=100\%$ with a $1\sigma$ one-sided lower-bound (64th percentile) at ${\sim}\,85\%$. p-values for the likelihood-ratio test (LRT) comparing the model with the corresponding constant model are included. Flux-weighted integrated polarization is reported for comparison with the Ep1 PWN \texttt{PCUBE} polarization in Section \ref{sec:pol}. Note that we have only recorded the most significant $\theta$-$\phi_0$ mode here. For comparison, the Ep2 best-fit constant model has $p_{0,neb}=9.7\pm1.8\%$ and $\psi_{0,neb}=35\pm5^\circ$.\label{tab:rvm_unbinned}}
\end{table*}

Since the statistical significance is limited and since, in reality, $p_{\rm psr}$ is likely phase dependent, our geometrical conclusions are preliminary. Our main purpose is to demonstrate that a strongly-polarized RVM sweep can plausibly explain the ${>}3\sigma$ difference in the Ep1/Ep2 PWN \texttt{PCUBE} polarization measurements. We can conclude that: (1) for Ep1, the RVM model is strongly favored over a constant polarization model; (2) an RVM-sweeping anomalous pulse is consistent with the low phase-integrated PWN polarization in Ep1; and (3) the RVM model seems to agree with the externally-modeled $\zeta\,{\sim}\,118^\circ$ and imply a large $\beta$, consistent with the lack of radio detection from the pulsar. The main pulse polarization is not significantly detected in the first or second IXPE observations. Additional data would help verify the pulsar geometry and potentially allow us to detect a more weakly polarized main pulse.

\section{Discussion} \label{sec:discussion}

We have conducted a detailed analysis of the polarization, timing, and spectral properties of Kes 75 during IXPE observations taken in Oct/Nov 2024 (Ep1) and April 2025 (Ep2). According to the polarization contour plot (Figure \ref{fig:pwn_pd}), the phase-integrated PSR/PWN's polarization differs by ${>}\,3\sigma$ between the two observations. The difference mainly arises in Stokes U, with $\Delta u = 0.09$ at a significance of $3.6\sigma$, corresponding to a statistical probability of $3.2 \times 10^{-4}$. Stokes Q is not significantly different between the two observations.

The PSR/PWN polarization is significant during Ep2, with PA $=36.8^\circ \pm7.3^\circ$ near the torus symmetry axis $\Psi=207^\circ(27^\circ)\,\pm\,8^\circ$, measured by \cite{Ng2008} from Chandra imaging. A similar PA $=25^\circ \pm 8^\circ$ is detected at a lower significance in the combined Ep1/Ep2 data. The EVPA implies a dominant toroidal magnetic field. Other PWNe with well-defined torus-jet structure also have IXPE EVPA aligned with the torus symmetry axis \citep{Xie2022,Bucciantini2023, Wong2024, Bucciantini2025,DiLalla2025}.  

Unusually, we find that the 2--4 GHz radio polarization ($\rm PD_{2-4\,GHz} = 14\%$) is higher than the X-ray polarization ($\rm PD_{2-8\,keV} = 10\%$, Ep2), which is unexpected if the synchrotron-cooled radio electrons occupy more complex spatial zones. However, the $3\sigma$ upper limit for Ep1 and Ep2 are $\rm PD=12\%$ and $\rm PD = 17\%$, respectively, so the low X-ray PD could be a statistical fluctuation. The real puzzle is why the PSR/PWN polarization is insignificant in the Ep1 data.

PSR J1846$-$0258's high dipole field and past outbursts of magnetar-like activity suggest that pulsar variability is a possible explanation. Our timing analysis indicates no dramatic change in the spin-down behavior. However, we do detect an anomalous pulsed component, occurring about $\Delta\phi=+0.5$ from the main component, in Ep1. Assuming the same spectrum, the extra pulsed component would contribute about $1/3$ of the main pulsed component flux, or $\mathcal{F}_{{\rm 2-8\,keV}}\,{\sim}\,0.6\times10^{-12}\rm\,erg\,cm^{-2}\, s^{-1}$, using the pulsed flux measurement by \cite{Gotthelf2021}, well within the IXPE measurement uncertainties (see Table \ref{tab:pcube}). This explains the lack of strong flux or spectral variations in the PSR/PWN complex between IXPE epochs. 

Moreover, if highly polarized, the extra component can provide sufficient polarized flux to push an RVM signal to a significant detection, while suppressing the average polarization such that the background PWN is no longer well detected. We can speculate that a magnetospheric change has caused an additional zone (plausibly from the anti-pole of the normal emission) to become active and that the zone emits highly polarized X-rays. 

High polarization has been detected previously by IXPE from magnetars. \cite{Zane2023} measured $\rm PD\,\sim\,80\%$ at 6--8 keV from AXP 1RXS J1708. They inferred a thermal atmospheric hotspot, which can produce high polarization due to the preferential escape of X-mode photons, up to ${\sim}\,70\%-80\%$ for particular viewing geometries \citep{Gonzalez2016, Taverna2020}. \cite{Rigoselli2025} and \cite{Stewart2025} independently measured $\rm PD \sim 40-65\%$ in a high-energy power-law component in 1E 1841-045 that may originate from synchrotron or curvature radiation. A measurement of the anomalous pulse spectrum would have been helpful to distinguish between such different emission mechanisms. 

Our best-fit RVMs have large $\beta$, which would explain the non-detection of radio pulsations \citep{Juntao2025}, even during the 2006 and 2020 outbursts \citep{Gavriil2008, Blumer2021}. A large $\beta$ for the main pulse implies a small impact angle for the opposite pole. If the opposite pole produces transient radio emission, its radio counterpart may be detectable during a future anomalous pulse epoch, even if not associated with an X-ray outburst. Contemporary radio and high-sensitivity X-ray monitoring observations could test this possibility. 

X-ray pulse profile changes are common from magnetars during outbursts \citep{Kaspi2017}, but rotation-powered pulsars are known to have very stable lightcurves. Magnetar-like PSR J1119-6127 and PSR J1846-0258 are two exceptions. During its 2016 outburst, PSR J1119-6127 developed pulsed emission above 2.5 keV \citep{Archibald2016}, which shows only an upper limit during quiescence \citep{Gonzalez2005,Ng2012}. PSR J1846-0258's profile did not apparently change during its 2006 outburst \citep{Kuiper2009} but during its 2020 outburst, its 2.5-10 keV pulsed amplitude increased and a small shoulder emerged on the pulse leading edge \citep{Hu2023}. 

In our analysis, we find evidence for a lightcurve change without the typical indications of an outburst (e.g. glitch, significant X-ray flux increase), though this may be due to insufficient coverage and/or limited sensitivity. Nevertheless, we received no alerts from Swift BAT, Fermi GBM, and other high-energy monitors, suggesting that small magnetospheric or crustal activity may exist that can modify the X-ray light curve without triggering the dramatic X-ray flux changes that have been the hallmark of its ``magnetar-like" behavior.

Future high-sensitivity observations could probe the extra pulse, but this state with a second, RVM-dominated peak may be rare. Thus, deciphering the complex polarization of this system may require much higher signal-to-noise and/or many visits to catch the pulsar in its anomalous state again.

\section{Conclusion}\label{sec:conclusion}
We have studied X-ray spectral, timing, and polarization properties of SNR Kes 75's PWN for two IXPE epochs. In the April 2025 epoch (Ep2), we detect 2--8 keV integrated polarization in the PSR/PWN complex with $\rm PD\,{\sim}\,10\%$ and PA roughly aligned with the torus symmetry axis. Its phase-resolved lightcurve is consistent with the historical profile. In contrast, during Oct/Nov 2024 epoch (Ep1), we find only an upper-limit on the polarization. The lightcurve is significantly different from the historical profile and better matches a two-pulsed profile, with a new component at phase offset $\Delta \phi \approx 0.5$. This component is faint, lying within the flux uncertainties, and insignificantly shifts the phase-averaged spectrum.

We modeled the Ep1 unbinned phase-resolved polarization with the rotating vector model (RVM), finding a best-fit model at 99.5\% significance that is consistent with the PWN morphology. It suggests a highly polarized extra pulsed component with a large impact parameter $\beta$, which would explain the non-detection of radio pulsations. Furthermore, the phase-integrated model polarization is consistent with the Ep1 PSR/PWN polarization. Our RVM fitting does not yet discriminate between synchrotron and curvature emission modes, and we do not attempt to identify an explanation for the change in the pulse profile. However, given its prior magnetar-like activity, magnetospheric changes in PSR J1846-0258 are not unexpected. Further X-ray spectral, timing, and polarization studies will help elucidate its complex behavior.


\begin{acknowledgments}
This work was supported in part by NASA grants 80NSSC25K0277 and 80NSSC25K7523. F.X. is supported by National Natural Science Foundation of China (grant No. 12422306 and No.12373041), and Bagui Scholars Program. S.Z. and C.-Y. N. are supported by a GRF grant of the Hong Kong Government under HKU 17304524. This paper employs a list of Chandra datasets, obtained by the Chandra X-ray Observatory, contained in the Chandra Data Collection ~\dataset[DOI: 10.25574/cdc.611]{https://doi.org/10.25574/cdc.611}.
\end{acknowledgments}

%
\facilities{CXO, IXPE, NICER}

\software{ixpeobssim \citep{Baldini2022}, leakagelib \citep{Dinsmore2025}}


\appendix
\section*{Solar Background Removal} \label{appendix:solar}

\begin{figure*}
\centering
\includegraphics[width=\linewidth]{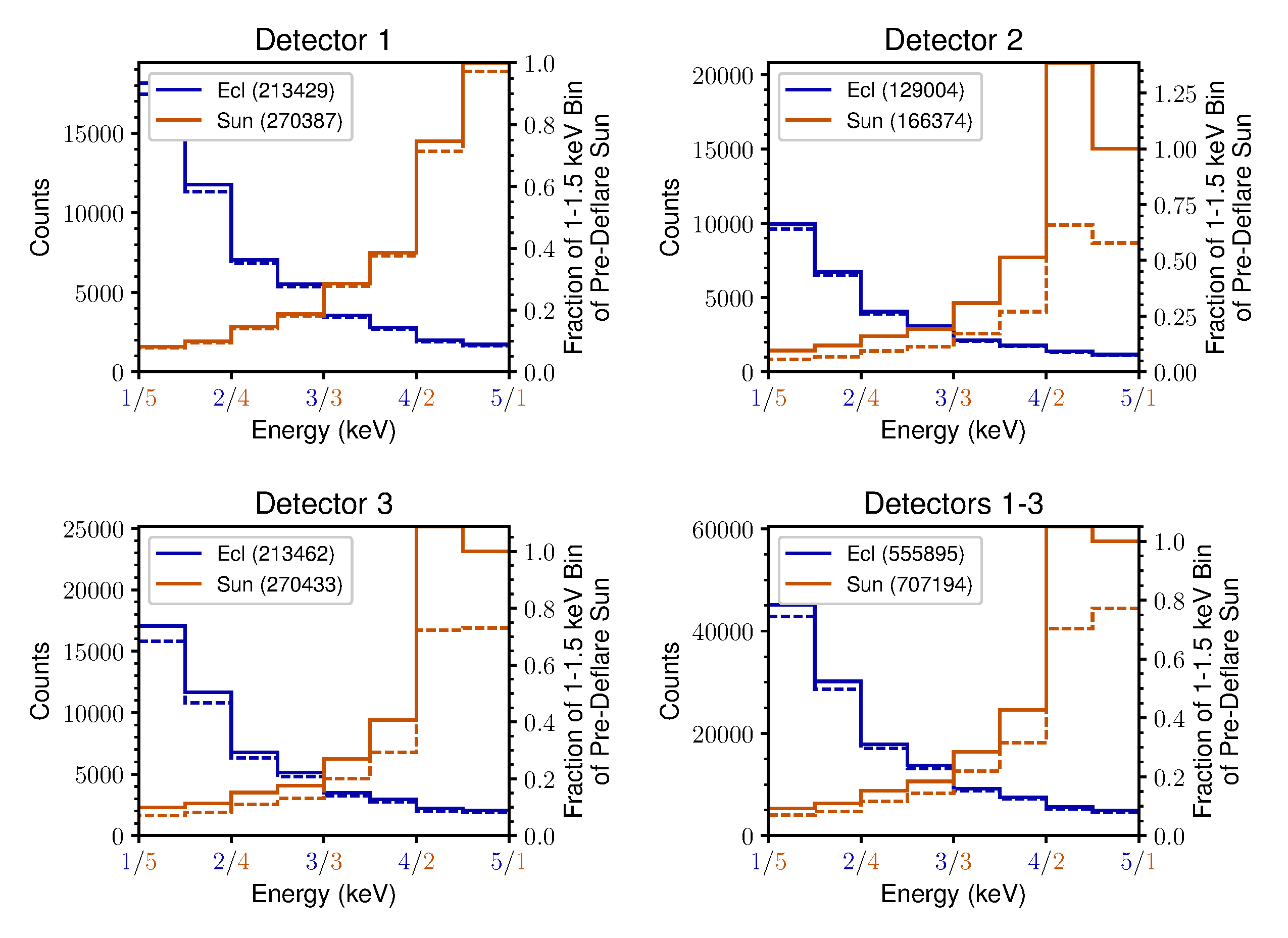}
\caption{IXPE 04002301 1--5 keV histograms before (solid) and after (dashed) deflaring, divided into in-eclipse (blue) and in-Sun (orange, reversed x-axis) epochs. Epoch livetimes (s) are listed in the legend. To facilitate comparison, the in-Sun histogram have been rescaled to the in-eclipse spectra using the livetime ratio. Left vertical axis represents bin counts, while the right axis normalizes to total counts in the 1--1.5 keV band, and the axes apply to both histograms.
After deflaring, the 1--2 keV bins dominate the In-Sun/Eclipse difference, and likely include residual solar photons, but this is outside the calibrated IXPE energy range and not used in the polarization analysis.}
\label{fig:deflare_spectrum}
\end{figure*}

To remove solar flares, which present a strongly polarized background, from the IXPE data, we utilized two different methods: (1) assigning time cuts during periods of strong flaring activity and (2) characterizing the spectral and polarization properties of the flares and subtracting their contributions.

\subsection*{Method 1: Time Cuts during Major Flares}

Utilizing the method described in \cite{Bucciantini2025}, we binned the cleaned data into 15-minute intervals and removed bins that exceeded $1.4\times$ ($1.5\times$ for moments-processed data) the median count rate across the entire observation. Each detector was processed individually since they have different levels of Sun exposure, depending on the spacecraft orientation. The thresholds were determined by comparing the lightcurves against solar flare data downloaded from the Geostationary Operational Environmental Satellite (GOES) database and identifying the value such that all M-class flares would be removed. Figure \ref{fig:deflare_spectrum} shows the event spectrum between $\rm 1-5\,keV$ before and after deflaring when the spacecraft is exposed to the sun versus in eclipse for the second IXPE observation. Deflaring removed proportionally more counts in DU2 and DU3 (${\sim} 20\%$) than in DU1 (${\sim} 5\%$), and the majority of these events occurred when the spacecraft was exposed to the Sun; the exact percentage varies by detector. Beyond 5 keV, solar photons have a negligible contribution to the spectrum. One can see that after deflaring the in-Sun spectrum more closely resembles the eclipse spectrum.

As noted in \cite{Bucciantini2025}, although deflaring helps decrease the solar background, most of its flux comes from low-level near-continuous emission. To account for this flux, as well as any unpolarized instrumental component, we selected an annular background region centered on the pulsar from $r=180''-280''$. This background region is close to that suggested by \cite{DiMarco2023}, adjusted for the extended nature of our target and the FOV boundary for our observation. This background subtraction removes both solar and particle background.

Even without this background subtraction, residual solar background after simple time cuts has a negligible effect on the polarization measurements within the PWN region. By assessing the in-sun and in-eclipse events after deflaring within the PWN region (defined as $r < 30''$ from the pulsar, see Section \ref{sec:pol}), we estimate that less than 1\% of the events between $\rm 2-8\,keV$ are residual solar photons. If we assume 100\% polarized solar photons with an orthogonal polarization angle to a 10\% polarized source (and no unpolarized background), the reduction of the polarization fraction would be less than 1\%. Indeed, the Ep2 PWN polarization, which is detected at a significant level (see Section \ref{sec:pol}), increases by ${\sim}1\%$ after background subtraction.

\subsection*{Method 2: Characterize and Subtract Solar Flares}
A simple time cut assumes no event energy dependence, while the in-Sun and in-eclipse spectra show that solar background dominates at low ($<2$ keV) energies. An alternative method characterizes the solar particle spectrum and polarization (and unpolarized instrumental background) and rescales to effect a background subtraction (Silvestri, in prep). We thus divided the event files into in-eclipse and in-Sun epochs and subtracted the in-eclipse data from the scaled in-Sun data to obtain the flare polarization and spectrum. For the instrumental background, we obtained stacked in-eclipse event energies and flux maps from two years of IXPE observations and ran a 1Ms \texttt{ixpeobssim} simulation to obtain an instrumental background template.

For the background subtraction, we re-scaled the flare and instrumental models to the exposure and the aperture of the source region and subtracted to obtain background-free polarization and spectra. Polarization results from this technique were consistent with those from simple de-flaring, but with reduced uncertainties. This improvement may be attributed to the higher signal-to-noise of the stacked instrumental background, and to the energy-dependent flare polarization.


\bibliographystyle{aasjournalv7}


\end{document}